\documentclass{IEEEtran}

\usepackage{algorithmic}
\usepackage{amsmath}
\usepackage{amssymb}
\usepackage{amsfonts}
\usepackage{epsfig}
\usepackage{setspace}
\usepackage{cite}
\usepackage{calc}
\usepackage{color}

\title{Explicit Codes Minimizing Repair Bandwidth for Distributed Storage}
\author{
Nihar B. Shah{\small $^\dagger$}, {K. V. Rashmi{\small $^\dagger$}, P. Vijay Kumar{\small $^\dagger$}, Kannan Ramchandran{\small $^\#$} }%
\vspace{1.6mm}\\
$^{^\dagger}$\,Dept. of ECE, Indian Institute Of Science, Bangalore. \{rashmikv, nihar, vijay\}@ece.iisc.ernet.in\\
$^{\#}$\,Dept. of EECS, University of California, Berkeley. kannanr@eecs.berkeley.edu\\
}

\newtheorem{thm}{Theorem}

\newtheorem{lem}[thm]{Lemma}
\newtheorem{cor}[thm]{Corollary}

\newcommand{\beq}{\begin{equation}}
\newcommand{\eeq}{\end{equation}}

\newcommand{\bea}{\begin{eqnarray}}
\newcommand{\eea}{\end{eqnarray}}
\newcommand{\bean}{\begin{eqnarray*}}
\newcommand{\eean}{\end{eqnarray*}}
\newcommand{\bit}{\begin{itemize}}
\newcommand{\eit}{\end{itemize}}
\newcommand{\ben}{\begin{enumerate}}
\newcommand{\een}{\end{enumerate}}
\newcommand{\blem}{\begin{lem}}
\newcommand{\elem}{\end{lem}}
\newcommand{\bthm}{\begin{thm}}
\newcommand{\ethm}{\end{thm}}
\newcommand{\bpf}{\begin{proof}}
\newcommand{\epf}{\end{proof}}

\begin{document}

\maketitle
\thispagestyle{empty}

\bibliographystyle{ieeetran}

\begin{abstract}
We consider the setting of data storage across $n$ nodes in a distributed manner. A data collector (DC) should be able to \textit{reconstruct} the entire data by connecting to any $k$ out of the $n$ nodes and downloading all the data stored in them. When a node fails, it has to be \textit{regenerated} back using the existing nodes. An obvious means of accomplishing this is to use a Reed-Solomon type MDS code where each node stores a single finite field symbol and where one downloads the entire file for regeneration of a failed node. However, storing vectors in place of symbols makes it easy to extract partial information from a node, and helps in reducing the amount of download required for regeneration of a failed node, termed as \textit{repair bandwidth}.

Recently, there has been additional interest in storing data in \textit{systematic} form as no post processing is required when the DC connects to the $k$ systematic nodes. On failure of a systematic node, there is a need to regenerate it back quickly and exactly due to their preferred status. Replacement of a failed node by an exact replica is termed \textit{exact regeneration}.

In this paper, we consider the problem of minimizing the repair bandwidth for exact regeneration of the systematic nodes. The file to be stored is of size $B$ and each node can store $\alpha = B/k$ units of data. A failed systematic node is regenerated by downloading $\beta$ units of data each from $d$ existing nodes. We give a lower bound for the repair bandwidth for exact regeneration of the systematic nodes which matches with the bound given by Wu et al. For $d \geq 2k-1$ we give an explicit code construction which achieves the lower bound on repair bandwidth when the existing $k-1$ systematic nodes participate in the regeneration. We show the existence and construction of codes that achieve the bound for $d \geq 2k-3$. Here we also establish the necessity of \textit{interference alignment}. We prove that the bound is not achievable for $d \leq 2k-4$ when $\beta=1$, except for the case when $\alpha=1$ for which any $[n,k]$ MDS code will trivially achieve the bound. We also give a coding scheme which can be used for any $d$ and $k$, which is optimal for $d \geq 2k-1$.
\end{abstract}

\section{Introduction}
\begin{figure}[t]
\hspace{20pt}
\includegraphics[width=0.6\textwidth]{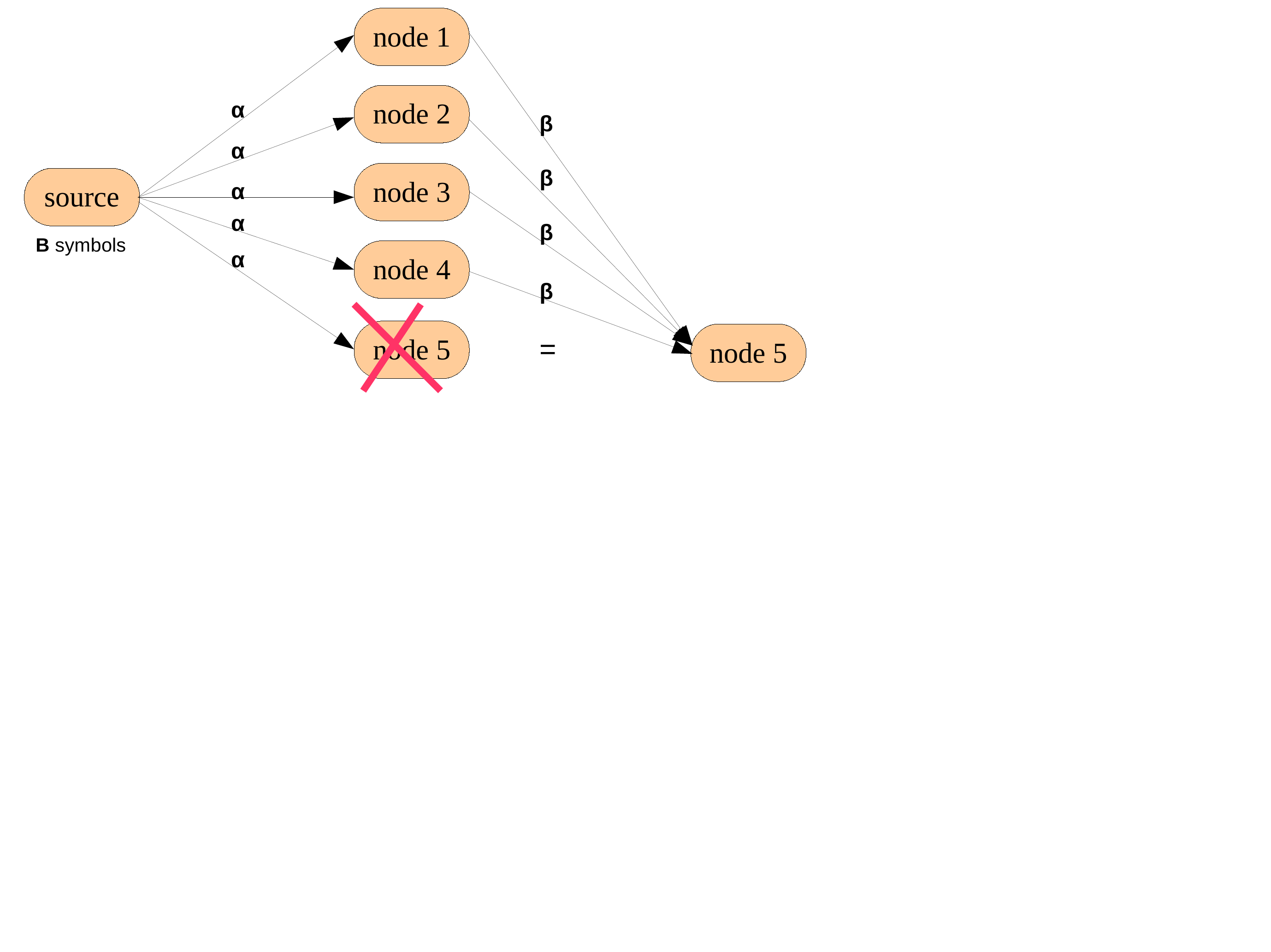}
\vspace{-150pt}
\caption{An illustration of exact regeneration: On failure of node 5, data from nodes 1 to 4 is used to regenerate back the same data that node 5 earlier had.} \label{fig:EgExact}
\vspace{-15pt}
\end{figure}

Consider a scenario where a file of size $B$ is to be stored in a distributed manner across $n$ storage nodes. A data collector (DC) should be able to \textit{reconstruct} the entire file by downloading data stored in any $k$ out of $n$ nodes. Each node can store $\alpha$ units of data (symbols) given by \beq \alpha=B/k \label{MSR_B} \eeq When a node fails, the failed node has to be \textit{regenerated} back by downloading $\beta$ symbols each from $d$ existing nodes as shown in Figure \ref{fig:EgExact}. We consider the problem of minimizing the repair bandwidth for exact regeneration of the systematic nodes.

Consider the exact regeneration of a systematic node, say node $l$ by connecting to some set of $d$ nodes. Each symbol stored is a linear function of the source symbols. By (\ref{MSR_B}) the linear functionals associated with the symbols stored in any $k-1$ of the $d$ nodes are linearly independent of those associated with the symbols of node $l$. Hence an additional $\alpha$ linear functionals are necessary to exactly regenerate node $l$. From this it follows that a lower bound on the repair bandwidth $d\beta$ is given by
\beq d\beta \geq \alpha + (k-1)\beta \label{MSR_d_general} \eeq 
In particular, for $\beta=1$ we have
\beq d \geq \alpha + k-1 \label{MSR_d}\eeq 

Our focus in the current paper is on the case $\beta=1$. Given a construction for $\beta=1$, constructions for larger $\beta$ can be obtained by partitioning the data into smaller chunks, and encoding them individually using the construction for $\beta=1$. As reconstruction and regeneration are performed separately on these smaller chunks, additional processing and storage required is greatly reduced. 
%Minimal bandwidth lower bound

\begin{figure}[t]
\hspace{-25pt}
\includegraphics[width=0.6\textwidth]{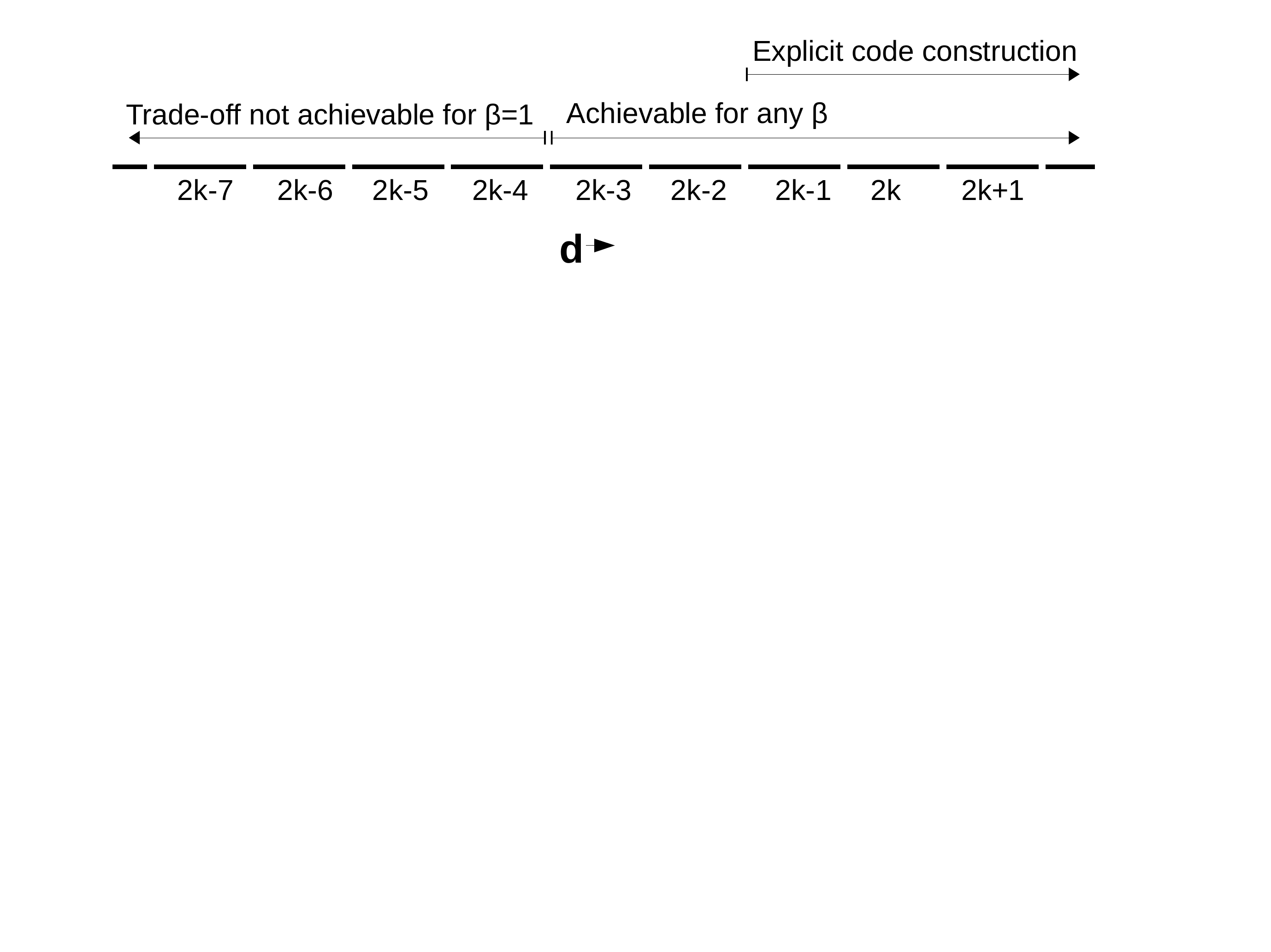}
\vspace{-185pt}
\caption{Categorization of the $(k,d)$ parameter set} \label{fig:organization}
\vspace{-15pt}
\end{figure}

In general, it is an open problem whether this lower bound is achievable for the problem of exact regeneration of the systematic nodes, and we address this issue in the present paper. For $\alpha=1$, we get $B=k$ and the lower bound as $d \geq k$. In this case, any $[n,k]$-MDS code will achieve the lower bound for exact regeneration. Hence, we will consider $\alpha > 1$ throughout. We have categorized the $(k,d)$ parameter set with respect to the lower bound on repair bandwidth in Figure \ref{fig:organization}.

In an independent work \cite{WuDimISIT}, authors consider the same setting and provide constructions for codes corresponding to a repair bandwidth that is significantly higher than the lower bound on repair bandwidth.

We say a code is \textit{optimal} exact regenerating if it achieves the the lower bound on repair bandwidth for the exact regeneration of systematic nodes. The non-systematic nodes are not the focus and hence their regeneration is not considered in detail. We assume the naive strategy of downloading the entire file for the exact regeneration of the non-systematic nodes.

The pioneering paper in this area \cite{YunDimKan} considers a more general setting in which each node stores slightly more data than the minimum required, namely $\alpha=(B/k)\nu$ for some $\nu \geq 1$, in order to reduce the repair bandwidth. In this scheme, the regenerated node need not be identical to the failed node as long as it maintains all the properties of the system. The authors establish a tradeoff between the amount of storage in each node $\alpha$ and the repair bandwidth $d\beta$. In the present paper, we are interested only in the case $\nu=1$ which corresponds to the Minimum Storage Regeneration (MSR) point on the tradeoff. The bound given in equation (\ref{MSR_d_general}) matches with the MSR point on the tradeoff.

In our previous work \cite{rasNih}, we also considered the problem of exact regeneration, however for that value of $\nu>1$ which minimized the repair bandwidth. We gave explicit codes for the other extreme point on the tradeoff, and an approximately exact regenerating code for the MSR point, both of which minimized the repair bandwidth at the respective points. 

In the rest of the paper, the results are presented in terms of $k$ and $\alpha$, as this leads to a more intuitive understanding of the codes. In Section \ref{sec:subspaceview} we give a subspace viewpoint which will be used throughout the paper. Explicit and optimal code constructions for $k \leq \alpha$ are given in Section \ref{sec:explicit}. The existence and construction of optimal codes for $k \leq \alpha + 2$ is given in Section \ref{sec:exist_alpha_2}. In Section \ref{sec:non_exist_alpha_3} we prove that the lower bound in not achievable for $k \geq \alpha+3$ with $\beta=1$. A coding scheme for any $(k,\alpha)$ parameter set is provided in Section \ref{sec:simple_ach} which is optimal for $k \leq \alpha$.

\section{Subspace Viewpoint for Linear Codes} \label{sec:subspaceview}

We consider only linear codes in this paper. By a linear code, we mean that any symbol stored is a linear combination of the source symbols, and only linear operations are allowed on them. Define a vector $\underline{\mathbf{z}}$ of length $B$ consisting of the source symbols. Let \beq \underline{\mathbf{z}}= \begin{bmatrix} \underline{z}_1 \\ \vdots \\ \underline{z}_k\end{bmatrix} \nonumber \eeq where $\underline{z}_i$ is a column vector of length $\alpha$. Each source symbol can independently take values from $\mathbb{F}_q$, a finite field of size $q$. Hence, the $B$ source symbols can be thought of as forming a $B$-dimensional vector space over $\mathbb{F}_q$.

Since the code is linear, any stored symbol can be written as $\underline{\ell}^t \underline{\mathbf{z}}$ for some column vector $\underline{\ell}$. These vectors which specify the kernels for the stored symbols define the code, and the actual symbols stored depend on the instantiation of $\underline{\mathbf{z}}$. Since a node stores $\alpha$ symbols, it can be considered as storing $\alpha$ vectors of the code, and hence can be represented by a $\alpha \times B$ matrix. We will say that the node \textit{stores} this matrix. 

Linear operations performed on the stored symbols are equivalent to the same operations performed on the corresponding vectors. Hence storing an $\alpha \times B$ matrix is equivalent to \textit{storing a subspace} of dimension at most $\alpha$. However, from (\ref{MSR_B}) it is clear that each node must store a subspace of dimension at least $\alpha$.

For $m=1,\ldots,n$ denote the matrix stored by node $m$ as $\mathbf{G}^{(m)} = [G^{(m)}_1 \quad G^{(m)}_2 \quad \ldots \quad G^{(m)}_k]$, where $G^{(m)}_l, l=1,\ldots,k$ are $\alpha \times \alpha$ matrices. The $\alpha$ symbols stored by  node $m$ are $\mathbf{G}^{(m)} \underline{\mathbf{z}} = \sum_{l=1}^{k} G^{(m)}_l \underline{z}_l$. We will denote the $j^{th}$ row of $G^{(m)}_l$ as $\underline{g}^{(m)}_{jl}$.

There are $n$ storage nodes, out of which $k$ are systematic and store $\alpha$ data symbols each in uncoded form. For $m=1,\ldots,k$, systematic node $m$ stores the symbol set $\underline{z}_m$. Thus for $l=1,\ldots,k$,
\bea
G^{(m)}_l =
\left \lbrace \begin{array}{ll}
         I_{\alpha}  &\text{if } l=m \\
0_\alpha &\text{if } l\neq m
        \end{array} \right.
\eea
where $0_\alpha$ is $\alpha \times \alpha$ zero matrix, and $I_\alpha$ is $\alpha \times \alpha$ identity matrix.  Hence, for any non-systematic node $m$, $G^{(m)}_l$ denotes the \textit{components} along systematic node $l$ that are stored in node $m$.

\begin{figure}[t]
%\vspace{-15pt}
\hspace{60pt}
\includegraphics[width=0.6\textwidth]{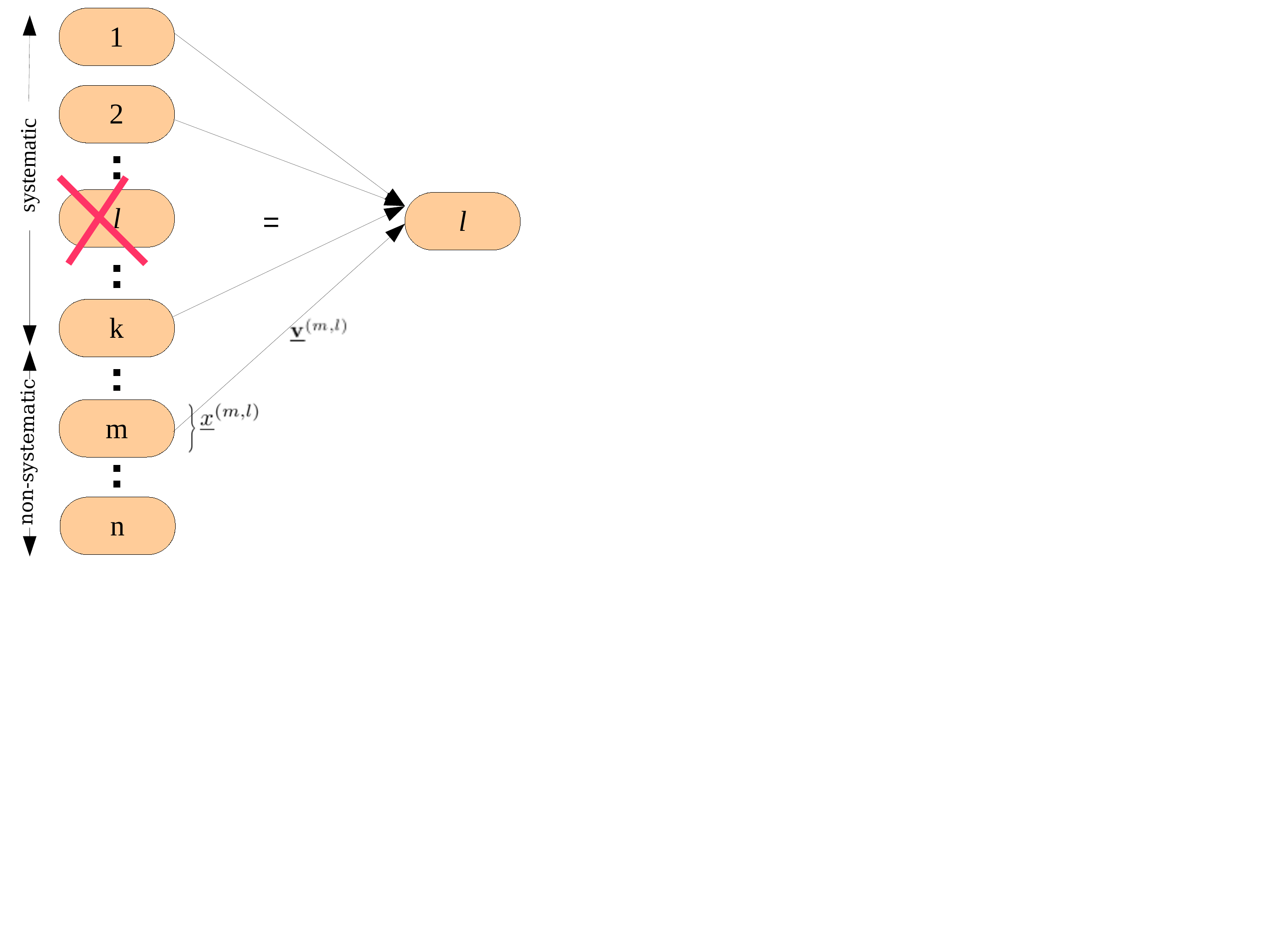}
\vspace{-95pt}
\caption{Exact regeneration of systematic node $l$} \label{fig:exact_regen}
\vspace{-20pt}
\end{figure}

For regeneration of a failed systematic node, $d$ other nodes provide one symbol each. We say that each node \textit{passes a vector} for the regeneration of the failed node. In the vectors passed by the non-systematic nodes, the components along the existing systematic nodes constitute \textit{interference}.

Let $\mathbb{D}$ denote the set of $d$ existing nodes used for regeneration of systematic node $l$.
Let $\underline{\mathbf{v}}^{(m,l)}_{\mathbb{D}}=[\underline{v}^{(m,l)}_{\mathbb{D}, 1} \cdots \underline{v}^{(m,l)}_{\mathbb{D}, k}]$ represent the vector passed by node $m \in \mathbb{D}$ for the regeneration of node $l \notin \mathbb{D}$ (as shown in Figure \ref{fig:exact_regen}) where $\underline{v}^{(m,l)}_{\mathbb{D}, i},\;(i=1,\ldots,k)$ is an $\alpha$-length row vector representing the component along the symbols of the systematic node $i$. Thus, $\underline{v}^{(m,l)}_{\mathbb{D}, i},\;(i=1,\ldots,k,\;\; i \neq l)$ constitute interference.
Let $\underline{x}^{(m,l)}_{\mathbb{D}}=[x^{(m,l)}_{\mathbb{D}, 1}\ldots\; x^{(m,l)}_{\mathbb{D}, \alpha}]$ be the coefficients of the linear combination of the rows of $\mathbf{G}^{(m)}$ to obtain the vector that node $m$ passes for regeneration of systematic node $l$. For brevity, we will discard the subscript $\mathbb{D}$ from the notation and the set of $d$ nodes being used for regeneration will be clear from the context. Thus, \beq \underline{\mathbf{v}}^{(m,l)} = \underline{x}^{(m,l)}\mathbf{G}^{(m)} \eeq 

Throughout this paper, we use superscripts to refer to the node numbers, and subscripts to index the elements of any matrix. No distinction is made between row and column vectors and the orientation of the vector under consideration is clear from the context. $\underline{e}_i$ represents an $\alpha$-length unit vector with $1$ in $i^{th}$ position and $0$ elsewhere. We say two vectors are \textit{aligned} if they are linearly dependent. 

\section{Optimal Explicit Code for $k \leq \alpha$} \label{sec:explicit}

In this section an explicit linear construction is given, which achieves optimal exact regeneration of systematic nodes for $k \leq \alpha$. The construction assumes that when a systematic node fails, the existing $k-1$ systematic nodes along with any $\alpha$ non-systematic nodes participate in the regeneration. First, we provide a code construction for $k = \alpha$. Codes for any $k < \alpha$ can be obtained by modifying the code for $k=\alpha$. Initially, a simple example is given to illustrate the code.

\subsection{Example}
Take $k=\alpha=3$. This gives $d=5$ and $B=9$. Thus each node stores a $3 \times 9$ matrix. Let $n=6$ and $q=7$.

Let the first three nodes be systematic. Hence,
\begin{align}
\mathbf{G}^{(1)}\; &= [I_3\;\; 0_3\;\; 0_3] \\
\mathbf{G}^{(2)}\; &= [0_3\;\; I_3\;\; 0_3] \\
\mathbf{G}^{(3)}\; &= [0_3\;\; 0_3\;\; I_3]
\end{align}

\noindent
Let ${\Psi}_3$ = \( \left[  \begin{array}{lll}
{\psi}_1^{(4)} & {\psi}_2^{(4)} & {\psi}_3^{(4)} \\
{\psi}_1^{(5)} & {\psi}_2^{(5)} & {\psi}_3^{(5)} \\	
{\psi}_1^{(6)} & {\psi}_2^{(6)} & {\psi}_3^{(6)}
\end{array} \right] \)
be a $3 \times 3$ Cauchy matrix \cite{cauchy}. Any submatrix of a Cauchy matrix is full rank.

The three non-systematic nodes store the matrices $\mathbf{G}^{(m)},\;\; m=4,5,6,$ given by
\[ \left[ \resizebox{8.5cm}{!}{\begin{tabular}{*{9}{c}}
$2{\psi}_1^{(m)} $&$ 2{\psi}_2^{(m)} $&$ 2{\psi}_3^{(m)} $&$ {\psi}_2^{(m)}  $&$ 0          $&$ 0          $&$ {\psi}_3^{(m)}  $&$ 0          $&$ 0 $\\
$0          $&$ {\psi}_1^{(m)}  $&$ 0          $&$ 2{\psi}_1^{(m)} $&$ 2{\psi}_2^{(m)} $&$ 2{\psi}_3^{(m)} $&$ 0          $&$ {\psi}_3^{(m)}  $&$ 0$  \\
$0          $&$ 0          $&$ {\psi}_1^{(m)}  $&$ 0          $&$ 0          $&$ {\psi}_2^{(m)}  $&$ 2{\psi}_1^{(m)} $&$ 2{\psi}_2^{(m)} $&$ 2{\psi}_3^{(m)}$ \\
\end{tabular}} \right] \]

\subsubsection{Regeneration}
For the regeneration of systematic node  $l\;(\in \{1,2,3\})$, each non-systematic node passes its $l^{th}$ row. The choice of the non-systematic node matrices is such that in the vectors passed for regeneration of a systematic node, components along the existing systematic nodes (which constitute interference) are aligned. For example, consider regeneration of systematic node $1$. Each non-systematic node passes its first row. First rows of $\mathbf{G}^{(m)},\;\; m=4,5,6$, have components along the systematic nodes (nodes $2$ and $3$) aligned in the direction $[1\;0\;0]$. Now, the second and third systematic nodes pass $[0\quad0\quad0\quad1\quad0\quad0\quad0\quad0\quad0]$, and $[0 \quad 0 \quad 0 \quad0\quad0\quad0\quad1\quad0\quad0]$ respectively and cancel out the interference leaving behind the matrix $[{\Psi}_3\quad 0_3\quad 0_3]$. Since ${\Psi}_3$ is invertible, systematic node $1$ can be exactly regenerated.

\subsubsection{Reconstruction}\label{sec:eg_recon}
To reconstruct the entire data, DC can connect to any three nodes. For reconstruction to be possible, the $9 \times 9$ matrix formed by juxtaposing the node matrices of these three nodes one below the other should be non-singular.

Reconstruction is trivially satisfied when the data collector connects to all the three systematic nodes. Suppose, the data collector connects to two systematic nodes and one non-systematic node. For example, suppose DC connects to nodes 2, 3 and 4. For reconstruction, we need the following matrix to be non-singular:
\beq A_1 = G^{(4)}_1=
\left[ \resizebox{3cm}{!}{\begin{tabular}{ccc}
$2{\psi}_1^{(4)} $&$ 2{\psi}_2^{(4)} $&$ 2{\psi}_3^{(4)} $\\
$0 $&$ {\psi}_2^{(4)} $&$ 0$\\
$0 $&$ 0$&$ {\psi}_3^{(4)} $
\end{tabular}} \right] \eeq
which is full rank since the elements of a Cauchy matrix are non-zero.

Consider the data collector connecting to one systematic node and two non systematic nodes. For example, suppose it connects to nodes $1,\;4,\;\text{and }5$. Since all symbols of node $1$ are available, $G_1^{(4)}$ and $G_1^{(5)}$ can be cancelled out. Hence for reconstruction to be possible the matrix $B_1$ given below must be full rank.
\begin{align*} B_1 = \begin{bmatrix}
            G_2^{(4)} & G_3^{(4)} \\
G_2^{(5)} & G_3^{(5)}
           \end{bmatrix}
\end{align*}

\textit{Claim:}  The matrix $B_1$ is full rank.

\begin{IEEEproof}
For $i=2,3,1$ (in this order), group the $i^{th}$ rows of the two non-systematic nodes together to give matrix 
\[ B_2 = 
\left[ \resizebox{5.6cm}{!}{\begin{tabular}{ccc|ccc}
$ 2{\psi}_1^{(4)} $&$ 2{\psi}_2^{(4)} $&$ 2{\psi}_3^{(4)} $&$ 0          $&$ {\psi}_3^{(4)}  $&$ 0$  \\
$ 2{\psi}_1^{(5)} $&$ 2{\psi}_2^{(5)} $&$ 2{\psi}_3^{(5)} $&$ 0          $&$ {\psi}_3^{(5)}  $&$ 0$  \\
\hline
$ 0          $&$ 0          $&$ {\psi}_2^{(4)}  $&$ 2{\psi}_1^{(4)} $&$ 2{\psi}_2^{(4)} $&$ 2{\psi}_3^{(4)}$ \\
$ 0          $&$ 0          $&$ {\psi}_2^{(5)}  $&$ 2{\psi}_1^{(5)} $&$ 2{\psi}_2^{(5)} $&$ 2{\psi}_3^{(5)}$\\
\hline
$ {\psi}_2^{(4)}  $&$ 0          $&$ 0          $&$ {\psi}_3^{(4)}  $&$ 0          $&$ 0 $\\
$ {\psi}_2^{(5)}  $&$ 0          $&$ 0          $&$ {\psi}_3^{(5)}  $&$ 0          $&$ 0 $
\end{tabular}} \right] \]

Let ${\Psi}_2$ = \( \left[  \begin{array}{cc}
{\psi}_2^{(4)} & {\psi}_3^{(4)} \\
{\psi}_2^{(5)} & {\psi}_3^{(5)}
\end{array} \right] \)

${\Psi}_2$ is a submatrix of the Cauchy matrix ${\Psi}_3$ and hence is invertible. Multiply the three groups of two rows each by ${\Psi}_2^{-1}$ to obtain
\begin{align} B_3 =  \left[ \resizebox{2.5cm}{!}{\begin{tabular}{ccc}
${\Psi}_2^{-1} $&$0_3$&$0_3$ \\
$ 0_3$&${\Psi}_2^{-1} $&$0_3$ \\
$ 0_3$&$0_3$&${\Psi}_2^{-1}$
\end{tabular}} \right] B_2 \\
=\left[ \resizebox{3cm}{!}{\begin{tabular}{ccc|ccc}
$\phi  $&$ 2  $&$ 0  $&$ 0  $&$ 0  $&$ 0  $\\
$\phi  $&$ 0  $&$ 2  $&$ 0  $&$ 1  $&$ 0  $\\
\hline
$0  $&$ 0  $&$ 1  $&$ \phi  $&$ 2  $&$ 0  $\\
$0  $&$ 0  $&$ 0  $&$ \phi  $&$ 0  $&$ 2  $\\
\hline
$1  $&$ 0  $&$ 0  $&$ 0  $&$ 0  $&$ 0  $\\
$0  $&$ 0  $&$ 0  $&$ 1  $&$ 0  $&$ 0  $
\end{tabular}} \right] 
\end{align}
where $\phi$ is some arbitrary value. Rows 1, 4, 5, 6 (and columns 1, 2, 4, 6) are linearly independent of all others (this includes all the columns containing $\phi$) and hence be eliminated to obtain,

\[ B_4 = 
\left[ \resizebox{1cm}{!}{\begin{tabular}{cc}
$2  $&$ 1  $\\
$1  $&$ 2  $
\end{tabular}} \right] \]
which is full rank.
\end{IEEEproof}

Now consider the case of DC connecting to three non-systematic nodes. Let $C_1$ be the matrix formed by juxtaposing the matrices stored in these three nodes one below the other.

\textit{Claim:} The matrix $C_1$ is full rank.

\begin{IEEEproof}
In $C_1$, group the $i^{th}\; (i=1,2,3)$ rows of all the three nodes together to obtain the matrix $C_2$. Thus,
\beq \hspace{-8cm} C_2 = \eeq
\[ \resizebox{!}{1.2cm}{\begin{tabular}{c}
grp1\\
\\
\\
grp2\\
\\
\\
grp3

\end{tabular}} \!\!\!\!\!\left[ \resizebox{!}{1.7cm}{\begin{tabular}{ccc|ccc|ccc}
$2{\psi}_1^{(4)} $&$ 2{\psi}_2^{(4)} $&$ 2{\psi}_3^{(4)} $&$ {\psi}_2^{(4)}  $&$ 0          $&$ 0          $&$ {\psi}_3^{(4)}  $&$ 0          $&$ 0 $\\
$2{\psi}_1^{(5)} $&$ 2{\psi}_2^{(5)} $&$ 2{\psi}_3^{(5)} $&$ {\psi}_2^{(5)}  $&$ 0          $&$ 0          $&$ {\psi}_3^{(5)}  $&$ 0          $&$ 0 $\\
$2{\psi}_1^{(6)} $&$ 2{\psi}_2^{(6)} $&$ 2{\psi}_3^{(6)} $&$ {\psi}_2^{(6)}  $&$ 0          $&$ 0          $&$ {\psi}_3^{(6)}  $&$ 0          $&$ 0 $\\
\hline
$0          $&$ {\psi}_1^{(4)}  $&$ 0          $&$ 2{\psi}_1^{(4)} $&$ 2{\psi}_2^{(4)} $&$ 2{\psi}_3^{(4)} $&$ 0          $&$ {\psi}_3^{(4)}  $&$ 0$  \\
$0          $&$ {\psi}_1^{(5)}  $&$ 0          $&$ 2{\psi}_1^{(5)} $&$ 2{\psi}_2^{(5)} $&$ 2{\psi}_3^{(5)} $&$ 0          $&$ {\psi}_3^{(5)}  $&$ 0$  \\
$0          $&$ {\psi}_1^{(6)}  $&$ 0          $&$ 2{\psi}_1^{(6)} $&$ 2{\psi}_2^{(6)} $&$ 2{\psi}_3^{(6)} $&$ 0          $&$ {\psi}_3^{(6)}  $&$ 0$  \\
\hline
$0          $&$ 0          $&$ {\psi}_1^{(4)}  $&$ 0          $&$ 0          $&$ {\psi}_2^{(4)}  $&$ 2{\psi}_1^{(4)} $&$ 2{\psi}_2^{(4)} $&$ 2{\psi}_3^{(4)}$ \\
$0          $&$ 0          $&$ {\psi}_1^{(5)}  $&$ 0          $&$ 0          $&$ {\psi}_2^{(5)}  $&$ 2{\psi}_1^{(5)} $&$ 2{\psi}_2^{(5)} $&$ 2{\psi}_3^{(5)}$ \\
$0          $&$ 0          $&$ {\psi}_1^{(6)}  $&$ 0          $&$ 0          $&$ {\psi}_2^{(6)}  $&$ 2{\psi}_1^{(6)} $&$ 2{\psi}_2^{(6)} $&$ 2{\psi}_3^{(6)}$
\end{tabular}} \right] \]

Multiply the 3 groups of 3 rows each by ${\Psi}_3^{-1}$ to get a matrix $C_3$ given by
\begin{align} C_3 &= \left[ \resizebox{2.6cm}{!}{\begin{tabular}{ccc}
          ${\Psi}_3^{-1} $&$ 0_3 $&$ 0_3 $\\ 
$0_3 $&$ {\Psi}_3^{-1} $&$ 0_3 $\\ 
$0_3 $&$ 0_3 $&$ {\Psi}_3^{-1} $
\end{tabular}} \right] C_2 \nonumber \\
&= \left[ \resizebox{4.6cm}{!}{\begin{tabular}{ccc|ccc|ccc}
$2  $&$ 0  $&$ 0  $&$ 0  $&$ 0  $&$ 0  $&$ 0  $&$ 0 $&$ 0  $\\
$0  $&$ 2  $&$ 0  $&$ 1  $&$ 0  $&$ 0  $&$ 0  $&$ 0 $&$ 0  $\\
$0  $&$ 0  $&$ 2  $&$ 0  $&$ 0  $&$ 0  $&$ 1  $&$ 0 $&$ 0  $\\
\hline
$0  $&$ 1  $&$ 0  $&$ 2  $&$ 0  $&$ 0  $&$ 0  $&$ 0 $&$ 0  $\\
$0  $&$ 0  $&$ 0  $&$ 0  $&$ 2  $&$ 0  $&$ 0  $&$ 0 $&$ 0  $\\
$0  $&$ 0  $&$ 0  $&$ 0  $&$ 0  $&$ 2  $&$ 0  $&$ 1 $&$ 0  $\\
\hline
$0  $&$ 0  $&$ 1  $&$ 0  $&$ 0  $&$ 0  $&$ 2  $&$ 0 $&$ 0  $\\
$0  $&$ 0  $&$ 0  $&$ 0  $&$ 0  $&$ 1  $&$ 0  $&$ 2 $&$ 0  $\\
$0  $&$ 0  $&$ 0  $&$ 0  $&$ 0  $&$ 0  $&$ 0  $&$ 0 $&$ 2  $
\end{tabular}} \right] \nonumber \end{align}

Rows $1$, $2$ and $3$ in the groups $1$, $2$ and $3$ respectively (i.e rows 1, 5 and 9) are clearly independent of all others (and so are the corresponding columns). The remaining $6 \times 6$ submatrix can be rearranged to get the following form:
\[ C_4 = 
 \left[ \resizebox{3cm}{!}{\begin{tabular}{cc|cc|cc}
$2  $&$ 1  $&$ 0  $&$ 0  $&$ 0  $&$ 0  $\\
$1  $&$ 2  $&$ 0  $&$ 0  $&$ 0  $&$ 0  $\\
\hline
$0  $&$ 0  $&$ 2  $&$ 1  $&$ 0  $&$ 0  $\\
$0  $&$ 0  $&$ 1  $&$ 2  $&$ 0  $&$ 0  $\\
\hline
$0  $&$ 0  $&$ 0  $&$ 0  $&$ 2  $&$ 1  $\\
$0  $&$ 0  $&$ 0  $&$ 0  $&$ 1  $&$ 2  $
\end{tabular}} \right] \]
This is a block diagonal matrix, and since $q = 7$, is full rank.
Thus the matrix $C_1$ is full rank.
\end{IEEEproof}

\subsection{Explicit General Code Construction for $k= \alpha$}
Let $\Psi$ be an $(n-k) \times \alpha$ Cauchy matrix \cite{cauchy} with elements drawn from $\mathbb{F}_q$. i.e,
\bea
\Psi=\begin{bmatrix}
 \underline{\psi}^{(k+1)} \\
\underline{\psi}^{(k+2)}\\
\vdots \\
\underline{\psi}^{(n)}
\end{bmatrix}
\eea
where $\underline{\psi}^{(i)}=[{\psi}^{(i)}_1\;\ldots\;{\psi}^{(i)}_\alpha],\;i=k+1,\ldots,n$ are $\alpha$-length row vectors. Any submatrix of a Cauchy matrix is full rank. The minimum field size required for the construction of this Cauchy matrix is: \beq q \geq \alpha + n - k \eeq
Note that since $n-k \geq \alpha \geq 2$, we will have $q \geq 4$.

For $m=k+1,\ldots, n,\;\;i,j=1,\ldots,\alpha$, set
\beq \label{eq:choose_g}
\underline{g}^{(m)}_{ij}= 
\left\lbrace \begin{array}{ll}
\epsilon \underline{\psi}^{(m)} &\text{if }i = j \\
{\psi}^{(m)}_j\underline{e}_i,\;\; &\text{if }i\neq j

\end{array}
\right.
\eeq
where $\epsilon$ is any arbitrary value such that $\epsilon \neq 0 $ and $\epsilon^2 \neq 1$. Note that there always exists such a value if $q \geq 4$. 

As illustrated in the example, this choice makes the interference in the vectors passed by non-systematic nodes for the regeneration of a failed systematic node \textit{aligned}. This enables the existing systematic nodes to cancel the interference by passing one symbol each.

\subsubsection{Regeneration}
Consider regeneration of systematic node $\hat{l} (\in \{1,\ldots,k\})$. All non-systematic nodes who participate in the regeneration pass their $\hat{l}^{th}$ row, i.e. if non-systematic node $m (\in \{k+1,\ldots,n\})$ participates in the regeneration, then it passes
\bea \underline{\mathbf{v}}^{(m,\hat{l})}=[\underline{g}^{(m)}_{\hat{l},1}\;\ldots\;\underline{g}^{(m)}_{\hat{l},k}]
\eea
The systematic node $l \;(l=1,\ldots,k,\;l\neq \hat{l})$ passes \bea \underline{\mathbf{v}}^{(l,\hat{l})}=[\underline{0}\;\ldots\;\underline{0}\;\;\underline{e}_{\hat{l}}\;\;\underline{0}\;\ldots\;\underline{0}] \eea with $\underline{e}_{\hat{l}}$ in the $l^{th}$ position. 

From equation (\ref{eq:choose_g}), $\underline{g}^{(m)}_{\hat{l},l}$ are all aligned along the direction of $\underline{e}_{\hat{l}}$. Hence $\underline{\mathbf{v}}^{(l,\hat{l})}$ can be used to remove interference along the systematic node $l$ from $\underline{\mathbf{v}}^{(m,\hat{l})},\;\forall m$ participating in the regeneration.

Also from equation (\ref{eq:choose_g}), $\underline{g}^{(m)}_{\hat{l},\hat{l}}$ are rows of the Cauchy matrix $\Psi$, and hence are linearly independent. Using these $\alpha$ linearly independent vectors, the systematic node $\hat{l}$ can be regenerated.

\subsubsection{Reconstruction} \label{sec:recon}
For reconstruction to be successful, the matrices stored in the $k$ nodes to which the DC connects, when juxtaposed one below the other, should form a $B \times B$ full rank matrix. Call this the \textit{reconstruction matrix} $R$. If the DC connects to the $k$ systematic nodes, then reconstruction is trivially satisfied. Consider DC connecting to $p$ non-systematic nodes, and $k-p$ systematic nodes, $1 \leq p \leq k$.  Let $\delta_1,\ldots,\delta_p$ be the $p$ non-systematic nodes to which DC connects and let $\Omega_1,\ldots,\Omega_p$ ($\Omega_1 <\ldots < \Omega_p$) be the $p$ systematic nodes to which DC does \textit{not} connect. 

Reconstruction is successful if and only if the $p\alpha \times p\alpha$ matrix $R$ formed by components along systematic nodes $\Omega_1,\ldots,\Omega_p$ in $\mathbf{G}^{(\delta_1)},\ldots,\mathbf{G}^{(\delta_p)}$ is non-singular.

\bea
R=\begin{bmatrix}
   \mathbf{G'}^{(\delta_1)} \\
\vdots \\
   \mathbf{G'}^{(\delta_p)}
  \end{bmatrix} = \begin{bmatrix}
G^{(\delta_1)}_{\Omega_1} & G^{(\delta_1)}_{\Omega_2} & \cdots & G^{(\delta_1)}_{\Omega_p} \\
\vdots & \vdots & & \vdots \\
G^{(\delta_p)}_{\Omega_1} & G^{(\delta_p)}_{\Omega_2} & \cdots & G^{(\delta_p)}_{\Omega_p}
\end{bmatrix}
\eea

\begin{thm} \label{thm:fullrank}
 $R$ is full rank.
\end{thm}

The proof of Theorem \ref{thm:fullrank} is provided in Appendix. The steps followed in the proof are similar to the ones used in the example.

\subsection{Explicit Code construction for $k < \alpha$}
For a given  $(k,\alpha)$ first construct the code for $k = \alpha$. The theorem given below shows the existence and construction for any $k < \alpha$. 

\begin{thm}
\label{thm:smaller_k}
If there exists a $(k,\alpha)$ linear code for exact regeneration of the systematic nodes, then there also exists a ($\hat{k},\alpha$) linear code for any $\hat{k} \leq k$.
\end{thm}

\begin{IEEEproof}
Suppose there exists an $(k,\alpha)$ code for exact regeneration of the systematic nodes. From each node matrix, remove the last $(k-\hat{k})\alpha$ columns so that now $\mathbf{G}^{(m)}$ is of the size $\alpha \times \hat{k}\alpha$. Thus, we will have $\hat{B} = \hat{k} \alpha$ data symbols. Consider only the set of first $\hat{k}$ systematic nodes and all the non-systematic nodes. This forms a ($\hat{k},\alpha$) code. 

\textit{Reconstruction:} Suppose the DC connects to $\ell$ systematic nodes and $\hat{k}-\ell$ non-systematic nodes. This case is same as the case in original $(k,\alpha)$ code where the DC connects to the removed $k-\hat{k}$ systematic nodes along with the above $\ell$ systematic nodes and $\hat{k}-\ell$ non-systematic nodes. Hence, the DC can reconstruct $\hat{B}$ data symbols. 

\textit{Regenration:} During regeneration of a systematic node, we have $\hat{d} = \alpha + \hat{k} - 1 = d - (k-\hat{k})$. All the $\alpha$ non-systematic nodes participating in the regeneration pass exactly the same vector as in the original code. Since the last $(k-\hat{k})\alpha$ column sets have been removed from the non-systematic node matrices, there is no interference from the data symbols corresponding to the removed $k-\hat{k}$ systematic nodes. All the interference which is present are aligned and the components along the failed systematic node span an $\alpha$-dimensional space as in the original code. Hence $k-\hat{k}$ lesser vectors will be able to regenerate the failed node. 
\end{IEEEproof}

\textit{Remark:} The above construction is optimal for $\beta = 1$. For any higher $\beta$, the data to be stored can be split into smaller chunks, which can be encoded individually using this construction for $\beta = 1$. Hence, this construction is optimal for any value of $\beta$.

\section{Existence and construction for $k \leq \alpha + 2$}\label{sec:exist_alpha_2}
The existence and construction of exact regenerating codes which meet the bound given by (\ref{MSR_d}) is shown for the parameter set $k \leq \alpha +2$. This proof assumes that when a systematic node fails, the existing $k-1$ systematic nodes participate in regeneration along with any $\alpha$ non-systematic nodes, passing one symbol each. The proof can be extended to the general case as well, where any $d$ existing nodes can participate in the regeneration.

\subsection{Approach}\label{sec:app}

In the sequel the reconstruction and regeneration conditions will be cast as product of rational polynomials. We will need to show that there exists a set of non-zero values such that these polynomials are all well defined and non-zero. In \cite{KotMed} a similar problem is arises in proving the existence of capacity achieving multicast network codes, but with respect to polynomials. But the argument can be easily extended to rational polynomials. If $\frac{f_1(\underline{x})}{g_1(\underline{x})},\ldots,\frac{f_p(\underline{x})}{g_p(\underline{x})}$ are rational polynomials, then define $f_{p+1}(\underline{x}) = gcd(g_1(\underline{x}),\ldots,g_p(\underline{x}))$. There exists a solution to $\underline{x}$ such that the product of the rational polynomials is well defined and non-zero if and only if there exists a solution to $\underline{x}$ such that the product of the polynomials $f_1(\underline{x}),\ldots,f_{p+1}(\underline{x})$ is non-zero. Hence, the algorithm given by Koetter and Medard in \cite{KotMed} can be used to find the values of the variables, provided the field size is large enough.

\subsection{Necessary Properties}

\subsubsection{Necessary Properties for Reconstruction}
\begin{lem}\label{lem:scalar} 
For reconstruction property to hold, for any non-systematic node $m$, $G_l^{(m)}$ must be full rank $\forall l \in \{1,\ldots,k\}$.
\end{lem}
\begin{IEEEproof}
Given some $m$ and $l$, suppose the DC connects to the $k-1$ systematic nodes other than $l$, and to the non-systematic node $m$. From the $k-1$ systematic nodes, the DC recovers $(k-1)\alpha$ data symbols. Hence column sets corresponding to these $k-1$ systematic nodes (i.e $G^{(m)}_{\hat{l}},\; \hat{l}=1,\ldots,k \;\; \hat{l} \neq l$) can be removed from $\mathbf{G}^{(m)}$ leaving behind only $G_{l}^{(m)}$. Thus, for successful reconstruction, $G_{l}^{(m)}$ should be full rank.\end{IEEEproof}

\subsubsection{Necessary Properties for Exact Regeneration}
\begin{lem}\label{lem:indep}
For the regeneration of a failed systematic node $l\;(\in \{1,\ldots,k\})$, the components along node $l$ in the vectors passed by the $\alpha$ non-systematic nodes participating in the regeneration must be linearly independent.
\end{lem}
\begin{IEEEproof}
Consider the regeneration of a failed systematic node $l$, by connecting to $k-1$ existing systematic nodes $m_1,\ldots,m_{k-1}$ and $\alpha$ non-systematic nodes $m_k,\ldots,m_{k-1+\alpha}$. Let matrix \beq \mathbf{V} = \left[ \begin{tabular}{c} $\underline{\mathbf{v}}^{(m_1,l)}$ \\ $\vdots$ \\$\mathbf{\underline{v}}^{(m_{k-1+\alpha,l})}$%\\$\mathbf{\underline{v}}_{k}$\\$\vdots$\\$\mathbf{\underline{v}}_{k-1+\alpha}$
\end{tabular} \right] = \left[ \begin{tabular}{cccc} $V_1$&$V_2$&$\cdots$&$V_k$ \end{tabular} \right] \eeq
where $V_i = $ \( \left[ \begin{tabular}{c} $\underline{v}^{(m_1,l)}_i$ \\ $\vdots$ \\$\underline{v}^{(m_{k-1+\alpha,l})}_i$\end{tabular} \right] \),$\quad (i=1,\ldots,k)$ is a $d \times \alpha$ matrix representing the component of $\mathbf{V}$ along the $i^{th}$ systematic node. For successful regeneration of $l$, we need an $\alpha \times d$ matrix $Y$ such that \beq Y\mathbf{V} = \mathbf{G}^{(l)} \eeq
Consider the component of $Y\mathbf{V}$ along node $l$. Since $G^{(l)}_{l} = I_{\alpha}$, we need $rank(YV_l) \geq \alpha$. Since the $k-1$ other systematic nodes cannot provide any vector in the direction of $l$, we get $\underline{v}_i^{(m,l)} = \underline{0}$ for $i,m=1,\ldots,k,i \neq m,\; l\neq m$. Thus the first $k-1$ rows of $V_l$ are $\underline{0}$. Hence, the remaining $\alpha$ rows of $V_l$, which are the components along the failed node in the vectors given by the non-systematic nodes, have to be linearly independent.\end{IEEEproof}
\textit{Remark:} Since only the last $\alpha$ rows of $V_l$ are non-zero, the last $\alpha$ columns of $Y$ should also be linearly independent.

\begin{lem} (Need for Interference Alignment)\label{lem:align}
For the regeneration of a failed systematic node $l$, and for any $\hat{l} \in 1,\ldots,k,\; \hat{l} \neq l$, the vectors $\underline{v}^{(m,l)}_{\hat{l}},\; \forall m \in \{k+1,\ldots,n\}$ should be aligned. \end{lem}
\begin{IEEEproof}
Using the same notations as in Lemma \ref{lem:indep}, consider the components along any other systematic node $\hat{l}$. Since $G_{\hat{l}}^{(l)}$ = $0_{\alpha}$, we need $YV_{\hat{l}}=0$.
Since the other $k-2$ systematic nodes provide $\underline{0}$ component along node $\hat{l}$, the corresponding rows of $V_{\hat{l}}$ will be zero. Let $\tilde{V}_{\hat{l}} \quad(\alpha+1 \times \alpha)$ and $\tilde{Y} \quad(\alpha \times \alpha+1)$ be sub-matrices of $V_{\hat{l}}$ and $Y$ with the $k-2$ zero rows in $V_{\hat{l}}$ and the corresponding columns in $Y$ removed. Thus we need $\tilde{Y} \tilde{V}_{\hat{l}} = 0$. Since $rank(\tilde{Y}) \geq \alpha$, it forces $rank(\tilde{V}_{\hat{l}}) \leq 1$. Hence, for the regeneration of a systematic node, in the vectors passed by the $\alpha$ non systematic nodes, the components along any existing systematic node should be aligned in the same direction. By choosing different sets of $\alpha$ non-systematic nodes, we get that alignment should hold for all the non-systematic nodes.\end{IEEEproof}
\begin{thm} \label{lem:sufficient}
A necessary and sufficient condition for exact regeneration of a failed systematic node $l$ by connecting to the existing $k-1$ systematic nodes and $\alpha$ non-systematic nodes is that the set of vectors passed by these non-systematic nodes  satisfy Lemmas \ref{lem:indep} and \ref{lem:align}.\end{thm}
\begin{IEEEproof}
\textit{Necessity}: Proved in Lemmas \ref{lem:indep} and \ref{lem:align} itself. \textit{Sufficiency}: Suppose Lemma \ref{lem:align} is satisfied. Then, in the vectors passed by the non-systematic nodes, the components along any other systematic node $\hat{l}$ are aligned in the same direction, i.e. $\underline{v}^{(m,l)}_{\hat{l}} = \kappa^{(m,l)}_{\hat{l}} \underline{w}^{(l)}_{\hat{l}}$ where $m$ is any non-systematic node, $\underline{w}^{(l)}_{\hat{l}}$ is a vector independent of $m$, and $\kappa$'s are some constants in $\mathbb{F}_q$. The systematic node $\hat{l}$ passes $\underline{v}^{(\hat{l},l)}_{\hat{l}} = \underline{w}^{(l)}_{\hat{l}}$ with the components of $\underline{\mathbf{v}}^{(\hat{l},l)}$ along other nodes as $\underline{0}$. Hence, this can be used to subtract the component in $\underline{\mathbf{v}}^{(m,l)}$ along any other systematic node $\hat{l}$, to give a set of vectors
\beq\underline{\tilde{\mathbf{v}}}^{(m,l)} = \underline{\mathbf{v}}^{(m,l)} - \sum_{i=1, i \neq l}^{k} \kappa^{(m,l)}_i \underline{\mathbf{v}}^{(i,l)}\eeq
Since $\underline{v}^{(\hat{l},l)}_l = \underline{0}\text{ for }\hat{l}=1,\ldots,k,\;\hat{l}\neq l$, we get $\underline{\tilde{v}}^{(m,l)}_l = \underline{v}^{(m,l)}_l$. Since Lemma \ref{lem:indep} is satisfied, the components of these $\alpha$ vectors along node $l$ are independent, and hence span the $\alpha$-dimensional subspace stored in node $l$.
\end{IEEEproof}

\subsection{Structure of the Code} \label{sec:codeform}

For $m=k+1,\ldots,n$ let,
\bea
G^{(m)}_i = \Lambda^{(m)}_i H^{(m)}_i,\quad i=1,\ldots,k
\eea
where $\Lambda^{(m)}_i=diag\{\lambda^{(m)}_{1,i},\ldots,\lambda^{(m)}_{\alpha,i}\}$ is an $\alpha \times \alpha$ diagonal matrix and
\bea
H^{(m)}_i=\begin{bmatrix}
           \underline{h}^{(m)}_{1,i} \\
\underline{h}^{(m)}_{2,i}\\
\vdots \\
\underline{h}^{(m)}_{\alpha,i}
          \end{bmatrix}
\eea
where $\underline{h}_{i,j}^{(m)}$ is an $\alpha$-length row vector.
Also set \beq \lambda^{(m)}_{i,i}=1, \quad i=1,\ldots,k \eeq

\textit{Regeneration:}
For $m=k+1,\ldots,n,\;\; l=1,\ldots,k$, let
\bea
\underline{\mathbf{v}}^{(m,l)}=\underline{x}^{(m,l)} \mathbf{G}^{(m)} \eea
For $l=1,\ldots,\alpha$, set \beq \underline{x}^{(m,l)}=\underline{e}_l \quad \forall m \in \{k+1,\ldots,n\} \eeq i.e. for regeneration of the systematic node $l (\in \{1,\ldots,\alpha\})$, each non-systematic node passes the $l^{th}$ row of its node matrix. Thus to satisfy Lemma \ref{lem:align} we choose,
\begin{align} \label{eq:theta_align}
\underline{h}^{(m)}_{i,j} = \underline{h}_{i,j}, \qquad m&=k+1,\ldots,n \\ 
i&=1,\ldots,\alpha, \nonumber \\ 
j&=1,\ldots,k, \quad j \neq i \nonumber \end{align}
Thus, for regeneration of systematic nodes $1,\ldots,\alpha$, the interference is aligned, and hence can be subtracted out.

\textit{Reconstruction:} The DC connects to any set of $k$ nodes and downloads all the $k\alpha$ data symbols stored in them.

\subsection{Existence and construction for $k=\alpha + 2$}
Consider regeneration of the systematic node $\alpha+1$. By Lemma \ref{lem:align}, the component along the systematic node $l,\;\forall l\in \{1,\ldots,\alpha\}$ in the vector passed by non-systematic nodes need to be aligned in one direction. This leads to the following set of $n-k-1$ equations: for $m=k+2,\ldots,n$,
\begin{align} \label{eq:set_one}
\kappa^{(m,\alpha+1)}_l \underline{x}^{(k+1,\alpha+1)} G^{(k+1)}_l = \underline{x}^{(m,\alpha+1)}G^{(m)}_l
\end{align}
Similarly, alignment for the regeneration of the systematic node $\alpha+2$ leads to another set of $n-k-1$ equations: $m=k+2,\ldots,n$,
\begin{align} \label{eq:set_two}
\kappa^{(m,\alpha+2)}_l \underline{x}^{(k+1,\alpha+2)} G^{(k+1)}_l = \underline{x}^{(m,\alpha+2)}G^{(m)}_l
\end{align}
for some constants $\kappa\text{'s} \in \mathbb{F}_q$. 

Set \begin{align} 
\kappa^{(m,\alpha+1)}_l&= \kappa^{(m,\alpha+2)}_l = \kappa^{(m)}_l \text{ (say) } \label{eq:ellkappa}
\end{align}

For all $m \in \{k+2,\ldots,n\}$, multiply equation (\ref{eq:set_one}) by $(x_l^{(m,\alpha+1)})^{-1}$ and (\ref{eq:set_two}) by $(x_l^{(m,\alpha+2)})^{-1}$ and subtract the two. $h_{l,l}^{(m)}$ gets eliminated and a homogeneous equation in terms of $\underline{h}_{1,l},\ldots,\underline{h}_{l-1,l},\;\underline{h}^{(k+1)}_{l,l},\;\underline{h}_{l+1,l},\ldots,\underline{h}_{\alpha,l}$ remains. One way to satisfy this equation is to equate all the scalar coefficients to zero.

This gives, for $l=1,\ldots,\alpha,\;m=k+2,\ldots,n$ and $i=1,\ldots,\alpha, i\neq l$,
\begin{align}
\lambda^{(m)}_{i,l} \!&=\!\kappa^{(m)}_l \!\lambda^{(k+\!1)}_{i,l}\!\! \left[ (x^{(m,\alpha+1)}_l)^{\!-\!1} x^{(k+1,\alpha+1)}_i - (x^{(m,\alpha+2)}_l)^{\!-\!1} \right.\nonumber \\
&\!\!\!\!\!\!\left. x^{(k+1,\alpha+2)}_i \right] 
\!\!\!\left[\! (x^{(m,\alpha+1)}_l)^{\!-\!1} x^{(m,\alpha+1)}_i \!\!\!-\! (x^{(m,\alpha+2)}_l)^{\!-\!1} x^{(m,\alpha+2)}_i \!\right]^{\!-\!1} \label{eq:am_a1}
\end{align}

Equation (\ref{eq:am_a1}) ensures that second set of equations (i.e. \ref{eq:set_two}) are satisfied whenever the first set (i.e. \ref{eq:set_one}) is satisfied. 

Note that any polynomial containing a $\lambda^{(m)}_{i,l} \quad (i \neq l)$ term will be a rational polynomial. For such polynomials, we will obtain an assignment which will simultaneously ensure that none of the inverted terms are zero, and the polynomial is also not zero.

Now only the first set of equations have to be satisfied, for which, using equation (\ref{eq:set_one}) make the following assignments, for $m=k+2,\ldots,n$
\begin{align}\label{eq:thetam}
\underline{h}^{(m)}_{l,l}&= (x^{(m,\alpha+1)}_l)^{-1} [ \kappa^{(m)}_l \{ \underline{h}^{(k+1)}_{l,l} x^{(k+1,\alpha+1)}_l + \nonumber\\ 
&\!\!\!\!\sum_{i=1, i \neq l}^{\alpha} \lambda_{i,l}^{(k+1)} x_i^{(k+1,\alpha+1)} \underline{h}_{i,l} \}\! -\!\!\! \sum_{i=1, i \neq l}^{\alpha}\lambda_{i,l}^{(m)} x_i^{(m,\alpha+1)} \underline{h}_{i,l} ]
\end{align}

The component along systematic node $\alpha+1$ needs to be aligned in the vector passed for the regeneration of systematic node $\alpha+2$ and vice versa. Hence the alignment of systematic nodes $\alpha+1$ and $\alpha+2$ result only in one set of $n-k-1$ equations each. Consider the regeneration of the $(\alpha+2)^{th}$ systematic node. By Lemma \ref{lem:align}, the the component along the $(\alpha+1)^{th}$ systematic node in the vector passed by non-systematic nodes need to be aligned in one direction. This leads to the following set of $n-k-1$ equations: For $m=k+2,\ldots,n$,
\begin{align} \kappa^{(m)}_{\alpha+1}\underline{x}^{(k+1,\alpha+2)}H_{\alpha+1}^{(k+1)}\Lambda_{\alpha+1}^{(k+1)} = \underline{x}^{(m,\alpha+2)}H_{\alpha+1}^{(m)}\Lambda_{\alpha+1}^{(m)}\end{align}
By equation (\ref{eq:theta_align}) we have \beq H_{\alpha+1}^{(m)} = H_{\alpha+1}^{(k+1)} = H_{\alpha+1} \text{ (say) } \eeq Thus, equating the coefficients to zero, we get for $i=1,\ldots,\alpha$,
\begin{align}\label{eq:a_alpha_1}
\lambda^{(m)}_{i,\alpha+1} = \kappa_{\alpha+1}^{(m)} x_i^{(k+1,\alpha+2)}\lambda^{(k+1)}_{i,\alpha+1}(x_i^{(m,\alpha+2)})^{-1} \end{align}
Similarly, for regeneration of node $\alpha+1$, we need to align components along node $\alpha+2$ which leads to
\begin{align}\label{eq:a_alpha_2}
\lambda^{(m)}_{i,\alpha+2} = \kappa_{\alpha+2}^{(m)} x_i^{(k+1,\alpha+1)}\lambda^{(k+1)}_{i,\alpha+2}(x_i^{(m,\alpha+1)})^{-1} \end{align}
\noindent\textit{Regeneration}: Exact regeneration of each one of the systematic nodes $l \in \{1,\ldots,\alpha\}$ results in a condition 
\beq \label{eq:sys1_det}
 det \left( \begin{array}{c}
      \underline{h}^{(m_1)}_{l,l} \\ \vdots \\ \underline{h}^{(m_\alpha)}_{l,l}
     \end{array} \right)
\neq 0
\eeq
where $m_1,\ldots,m_\alpha$ are the $\alpha$ non-systematic nodes used for regeneration
After substituting for $\underline{h}^{(m_i)}_{l,l},\quad i=1,\ldots,\alpha$ from equation (\ref{eq:thetam}), this condition evaluates to a rational polynomial, which can be shown to be not identically equal to zero by the following assignments: 
\begin{align}
&\kappa^{(m)}_l = 1,\quad \lambda^{(k+1)}_{i,l} = 1,\quad \underline{h}^{(k+1)}_{l,l} = \underline{e}_l,\quad \underline{h}_{i,l} = \underline{e}_i,\nonumber\\
&x_i^{(k+1,\alpha+1)}=0,\quad x_i^{(k+1,\alpha+2)}=1,\quad x_l^{(k+1,\alpha+1)}=1\nonumber\\
&x_l^{(m,\alpha+1)} = 1,\quad x_l^{(m,\alpha+2)} = 1, \quad x_i^{(m,\alpha+1)}=1 \nonumber\\
&x^{(m,\alpha+2)}_i = (m-k)^{-j} +1
\end{align}
for $i=1,\ldots,\alpha,\;i \neq l,\;m\in \{m_1,\ldots,m_\alpha\},m \neq k+1$ and $j=i\; \text{if}\; i<l, \; j=i-1 \;\text{if}\; i>l$.
This set of assignments makes the matrix under consideration in equation (\ref{eq:sys1_det}) a Vandermonde matrix which is full rank, and ensures that equations (\ref{eq:am_a1}), (\ref{eq:a_alpha_1}) and (\ref{eq:a_alpha_2}) remain valid, provided the field size is large enough.

Exact regeneration of systematic nodes $\alpha+1 \text{ and } \alpha+2$ also result in conditions of rational polynomials being not equal to zero. For exact regeneration of $(\alpha+1)^{th}$ systematic node, Lemma \ref{lem:indep} should hold. Choose $H_{\alpha+1}$ to be a full rank matrix. Lemma \ref{lem:indep} implies that the coefficients resulting from the linear combinations need to be linearly independent, i.e $\underline{x}^{(m,\alpha+1)}\Lambda^{(m)}_{\alpha+1}$ should be linearly independent for any $\alpha$ out of the $n-k$ non-systematic nodes. Express the determinant of this matrix as a polynomial. To show that this polynomial is not identically zero, we choose $\Lambda^{(k+1)}_{\alpha+1} = I, \kappa^{(m)}_{\alpha+1}=1, x^{(m,\alpha+2)}_i = 1, x^{(k+1,\alpha+2)}_i=1$ for $i=1,\ldots,\alpha,m=k+2,\ldots,n$.

From (\ref{eq:a_alpha_1}), we get $\Lambda^{(m)}_{\alpha+1} = I$.
Choose $x^{(m,\alpha+1)}_i = (m-k)^i$ for $m=k+1,\ldots,n$ to make it a Vandermonde matrix (also ensuring that equations (\ref{eq:am_a1}), (\ref{eq:a_alpha_1}) and (\ref{eq:a_alpha_2}) remain valid), provided the field size is large enough. A similar argument can be used to obtain a condition for regeneration of node $\alpha+2$.

\textit{Reconstruction}: The condition for reconstruction property to hold can be expressed as a product of polynomials not being equal to zero by viewing each determinant as a polynomial. For reconstruction to be successful, the node matrices corresponding to the $k$ nodes to which the data collector connects, when juxtaposed one below the other, should form a $B \times B$ full rank matrix. If the data collector connects to the $k$ systematic nodes, then reconstruction is trivially satisfied. Consider DC connecting to $p$ non-systematic nodes and $k-p$ systematic nodes, $1\leq p \leq k$. Let $m_1,\ldots,m_p,\quad(m_1<\ldots<m_p)$ be the non-systematic nodes to which it connects. Let $l_1,\ldots,l_p,\quad(l_1<\ldots<l_p)$ be the $p$ systematic nodes to which it does \textit{not}  connect. Due to the structure of node matrices of the systematic nodes, we will be left with the condition of the $l\alpha \times l\alpha$ matrix formed by the column sets $l_1,\ldots,l_p$ of the node matrices of the $p$ non-systematic nodes being non-singular. Thus the polynomial corresponding to this choice of $k$ nodes is
\beq det \left(
\resizebox{3.5cm}{!}{\begin{tabular}{cccc}
$G_{l_1}^{(m_1)} $&$ G_{l_2}^{(m_1)} $&$ \cdots $&$ G_{l_p}^{(m_1)}$ \\
$G_{l_1}^{(m_2)} $&$ G_{l_2}^{(m_2)} $&$ \cdots $&$ G_{l_p}^{(m_2)}$ \\
$\vdots          $&$ \vdots          $&$        $&$ \vdots         $ \\
$G_{l_1}^{(m_p)} $&$ G_{l_2}^{(m_p)} $&$ \cdots $&$ G_{l_p}^{(m_p)}$
\end{tabular}} \right)
\eeq

We will now show that there exists an assignment of the variables such that this polynomial is not identically zero. 
For $i,j=1,\ldots,p,\;m_1 \neq k+1$, set 
\beq \kappa^{(m_i)}_{l_j} = \left\lbrace \begin{array}{cc} 1 &\text{ if } i=j\\0 &\text{ if } i \neq j \end{array} \right. \eeq

For $i=1,\ldots,\alpha, m=k+2,\ldots,n, l=1,\ldots,k$ set
$H^{(k+1)}_l = I,\quad \Lambda^{(k+1)}_l = I, \quad x_i^{(m,\alpha+1)} = 1,\quad x_i^{(k+1,\alpha+1)} = 1,\quad x_i^{(k+1,\alpha+2)} = 1,\quad x_i^{(m,\alpha+2)} = (m-k)^i$.
Thus from (\ref{eq:am_a1}), (\ref{eq:thetam}), (\ref{eq:a_alpha_1}) and (\ref{eq:a_alpha_2}), we get that this matrix is full rank for a large enough field size, and also the equations (\ref{eq:am_a1}), (\ref{eq:a_alpha_1}) and (\ref{eq:a_alpha_2}) remain valid.

Hence, provided that the field size is large enough, one can find solutions for these variables such that both reconstruction and exact regeneration properties are satisfied.

\subsection{Existence and construction for $k < \alpha + 2$}\label{sec:exist_alpha_1}
For a given $\alpha$, the code described for the previous subsection can by modified using Theorem \ref{thm:smaller_k} to obtain a code for any $k < \alpha+2$.

\textit{Remark:} This achievability scheme is optimal for $\beta = 1$. For any higher $\beta$, the data to be stored can be split into smaller chunks, which can be encoded individually using this construction for $\beta = 1$. Hence, this for any value of $\beta$, optimal regeneration for the parameter set $k \leq \alpha+2$ is achievable using this scheme.

\section{Non-Achievability for $k \geq \alpha + 3$}\label{sec:non_exist_alpha_3}
We define two codes to be \textit{equivalent} if the corresponding nodes of both codes store the same subspace, and pass the same vectors for regeneration of any node. The only difference may be in the representation of what is stored in a node, i.e. in the node matrices.
\begin{lem} \label{lem:unique_struct}
If there exists an exact regenerating code for $k \geq \alpha+1$, then there exists an equivalent code with the following property:
\beq
\underline{h}^{(m)}_{i,j} = \underline{h}_{i,j}, \text{for } i=1,\ldots,\alpha,\;j=1,\ldots,k,\;,j\neq i
\eeq
for any non-systematic node $m$.
\end{lem}

\begin{IEEEproof} Consider a code which performs exact regeneration of the systematic nodes. For any non-systematic node $m$ in this code, we will obtain linearly independent vectors in the subspace stored in it one by one, and set them as the rows of its node matrix. By induction we will prove that the matrix of any node will take the following form:
$\mathbf{G}^{(m)} =$\beq \hspace{-.1cm}
\left[\hspace{-.15cm} \resizebox{8cm}{!}{ \begin{tabular}{*{6}{c}}
${\lambda}^{(m)}_{1,1}\underline{h}^{(m)}_{1,1} $&$\cdots$&$ {\lambda}^{(m)}_{1,\alpha}\underline{h}_{1,\alpha} $&$ {\lambda}^{(m)}_{1,\alpha+1}\underline{h}_{1,\alpha+1} $&$ \cdots $&$ {\lambda}^{(m)}_{1,k}\underline{h}_{1,k} $\\
${\lambda}^{(m)}_{2,1}\underline{h}_{2,1} $&$ \cdots $&$ {\lambda}^{(m)}_{2,\alpha}\underline{h}_{2,\alpha} $&$ {\lambda}^{(m)}_{2,\alpha+1}\underline{h}_{2,\alpha+1} $&$ \cdots $&$ {\lambda}^{(m)}_{2,k}\underline{h}_{2,k} $\\
$\vdots $&&$ \vdots $&&$ \vdots $\\
${\lambda}^{(m)}_{\alpha,1}\underline{h}_{\alpha,1} $&$ \cdots $&$ {\lambda}^{(m)}_{\alpha,\alpha}\underline{h}^{(m)}_{\alpha,\alpha} $&$ {\lambda}^{(m)}_{\alpha,\alpha+1}\underline{h}_{\alpha,\alpha+1} $&$ \cdots $&$ {\lambda}^{(m)}_{\alpha,k}\underline{h}_{\alpha,k}$
\end{tabular}} \label{eq:matrix}\hspace{-.15cm} \right]\hspace{-.15cm}\eeq
for $m=k+1,\ldots,n$. 
Note that by Lemma \ref{lem:scalar}, $H_l^{(m)}$ and $\Lambda_l^{(m)}$ have to be full rank matrices.

Let the first row of $\mathbf{G}^{(m)}$ represent the vector passed by the non-systematic node $m$ for the regeneration of the first systematic node. From Lemma \ref{lem:align}, as the interference from the remaining $k-1$ systematic nodes has to be aligned, \bea
\underline{h}^{(m)}_{1,j}= \underline{h}_{1,j},\;\; j=2,\ldots,k
\eea

Hence, the first row has to be of the given form. Suppose the vectors passed for the regeneration of systematic nodes $1,\ldots,p-1\;(1<p \leq \alpha) $ are linearly independent in all the non-systematic nodes. By a similar argument, the first $p-1$ rows of $\mathbf{G}^{(m)}$ have to be of the given form.

Now consider the regeneration of the $p^{th}$ systematic node. Suppose some of the non-systematic nodes (say type A) pass a vector linearly dependent on the first $p-1$ rows of their node matrix and some (say type B) pass a linearly independent vector. Each type B node will have this vector as a new row in their node matrices. In this set of vectors passed for regeneration of the $p^{th}$ systematic node, consider the component along systematic node $(\alpha+1)$, i.e. $\underline{v}_{\alpha+1}^{(m,p)}$. In vectors passed by type A nodes, this vector is a linear combination of $\underline{h}_{i,\alpha+1},\;\;i=1,\ldots,p-1$, whereas in vectors passed by type B nodes, it is linearly independent of $\underline{h}_{i,\alpha+1},\;\;i=1,\ldots,p-1$ (by Lemma \ref{lem:indep}). But by Lemma \ref{lem:align}, these vectors have to be aligned. Hence there is a contradiction.

Suppose all the non-systematic nodes are of type A. Then all the vectors passed by non-systematic nodes will be linearly dependent on the first $p-1$ rows of their node matrices. Hence in all the vectors, the component along systematic node p $\underline{v}^{(m,p)}_p$ will be a linear combination of $\underline{h}_{i,p},\;\;i=1,\ldots,p-1$. These can span at most $p-1$ dimensions whereas $p^{th}$ systematic node spans $\alpha$ dimensions. Hence the regeneration of $p^{th}$ systematic node is not possible.

Hence all the non-systematic nodes should be of type B. i.e they pass linearly independent vectors for the generation of systematic nodes. Along with Lemma \ref{lem:align}, this proves that all $\alpha$ rows of the node matrix have to be of the given form.
\end{IEEEproof}

Henceforth in this section, we will consider all nodes to be of this form.

\textit{Remark:} In this code, for the regeneration of the $p^{th}$ systematic node (for $1 \leq p \leq \alpha$), each non-systematic node passes the $p^{th}$ row of its node matrix. 
\begin{cor}\label{cor:HsameForOtherNodes}
For $l = \alpha+1,\ldots,k$, and any non-systematic nodes $m$ and $m^{\prime}$, \beq H^{(m)}_{l} = H^{(m^{\prime})}_{l} \eeq
\end{cor}

\begin{cor}\label{cor:vectorsPassedInd}
For any non-systematic node $m$ and $k \geq \alpha+1$, any $\alpha$ out of  $\underline{\mathbf{v}}^{(m,1)},\ldots,\underline{\mathbf{v}}^{(m,k)}$ are linearly independent.
\end{cor}
\begin{IEEEproof}
In the code given in Lemma \ref{lem:unique_struct}, the choice of the first $\alpha$ systematic nodes was arbitrary. Hence, for a given set of $\alpha$ systematic nodes, an equivalent code can be constructed considering these as the first $\alpha$ nodes. Thus, by Lemma \ref{lem:unique_struct} the vectors passed by any non-systematic node for regeneration of these systematic nodes will be the $\alpha$ rows of its matrix. Hence they are independent.
\end{IEEEproof}

\begin{lem} \label{lem:non_zero_coeffs}
For $k \geq \alpha + 2$, for any non-systematic node $m$, 
\begin{align} x_i^{(m,l)} \neq 0 \quad \text{ for } l\!=\!\alpha\!+\!1,\ldots,k,\; i=1,\ldots,\alpha \nonumber \end{align} 
\end{lem}

\begin{IEEEproof}Suppose for some non-systematic node $m$, $l \in \{\alpha\!+\!1,\ldots,k\}$ and $i\in \{1,\ldots,\alpha\}$, $x_i^{(m,l)}$ = $0$. Since for $j (\in \{1,\ldots,\alpha$\}), $\underline{\mathbf{v}}^{(m,j)}$ is the $j^{th}$ row of the node matrix of node $m$, $\underline{\mathbf{v}}^{(m,l)}$ is a linear combination of $\underline{\mathbf{v}}^{(m,j)},\;j=1,\ldots,\alpha,\;j \neq i$. This is a contradiction to Corollary \ref{cor:vectorsPassedInd}.
% 
% (By contradiction)Consider the regeneration of the systematic node $l$ for some $l \in \{\alpha+1,\ldots,k \}$. Suppose $x^{(m,l)}_i=0$ for some non-systematic node $m$ and some $l \in \{1,\ldots,\alpha \}$. 
% 
% For any systematic node $\hat{l} \in \{\alpha+1,\ldots,k\},\;\hat{l} \neq l$, the components along $\hat{l}$: $\underline{v}^{(m,l)}_{\hat{l}}\;(m=k+1,\ldots,n)$ need to be aligned in one direction(by Lemma \ref{lem:align}). Since the set of vectors $\underline{h}_{1,\hat{l}},\ldots,\underline{h}_{\alpha,\hat{l}}$ are linearly independent, we get $x^{(m,l)}_i=0\quad \forall m \in\{k+1,\ldots,n\}$.
% 
% Now consider the component along the failed systematic node $l$. Since $x^{(m,l)}_i=0\quad \forall m$, $\underline{v}^{(m,l)}_l\;\; \forall m$ are linear combinations of $\underline{h}_{j,l}\;(j=1,\ldots,\alpha,\; l \neq \hat{l}$) which can span only $\alpha-1$ dimensions. Thus all $\alpha$ dimensions of the systematic node $l$ cannot be regenerated.
\end{IEEEproof}

\begin{thm} \label{thm:non_exist}
For a linear code with $\beta=1$, exact regeneration of systematic nodes meeting the bound given by (\ref{MSR_d}) is not possible for $k \geq \alpha+3$.
\end{thm}

\begin{IEEEproof}
(By contradiction) Consider any code that achieves exact regeneration of systematic nodes meeting the bound in (\ref{MSR_d}) for $k \geq \alpha+3$. By Corollary \ref{cor:HsameForOtherNodes}, \beq \label{eq:Hnonsyssame} H^{(m)}_i = H^{(m^{\prime})}_i \eeq for $m,m^{\prime} \in k\!+\!1,\ldots,n,\; i=\alpha\!+\!1,\ldots,k$. Call these matrices $H_i$. 

Consider regeneration of systematic node $(\alpha+3)$. By Lemma \ref{lem:align}, components corresponding to systematic nodes $(\alpha+1)$ and $(\alpha+2)$ are to be aligned. Hence we have 
\bea
\underline{x}^{(k+1,\alpha+3)}G^{(k+1)}_{\alpha+1} &=& \kappa_1 \underline{x}^{(k+2,\alpha+3)}G^{(k+2)}_{\alpha+1}\\
\underline{x}^{(k+1,\alpha+3)}G^{(k+1)}_{\alpha+2} &=& \kappa_2 \underline{x}^{(k+2,\alpha+3)}G^{(k+2)}_{\alpha+2}
\eea
where $\kappa_1$ and $\kappa_2$ are some constants in $\mathbb{F}_q$.
From equation (\ref{eq:Hnonsyssame}) and since $H_{\alpha+1}$ is full rank (Lemma \ref{lem:scalar}), this simplifies to
\bea
\underline{x}^{(k+1,\alpha+3)}\Lambda^{(k+1)}_{\alpha+1} &=& \kappa_1 \underline{x}^{(k+2,\alpha+3)}\Lambda^{(k+2)}_{\alpha+1} \label{eq:alpha3_xSubstitute}\\
\underline{x}^{(k+1,\alpha+3)}\Lambda^{(k+1)}_{\alpha+2} &=& \kappa_2 \underline{x}^{(k+2,\alpha+3)}\Lambda^{(k+2)}_{\alpha+2}
\eea
\begin{align}
\implies \kappa_1 \underline{x}^{(k+2,\alpha+3)}\Lambda^{(k+2)}_{\alpha+1} (\Lambda^{(k+1)}_{\alpha+1})^{-1} & = & \nonumber\\
\kappa_2 \underline{x}^{(k+2,\alpha+3)}\Lambda^{(k+2)}_{\alpha+2} &(\Lambda^{(k+1)}_{\alpha+2})^{-1}&
\end{align}

Since no element of $\underline{x}^{(k+2,\alpha+3)}$ is zero (Lemma \ref{lem:non_zero_coeffs}), and the $\Lambda$ matrices are diagonal, we get
\bea
\kappa_1 \Lambda^{(k+2)}_{\alpha+1} (\Lambda^{(k+1)}_{\alpha+1})^{-1} & = & \kappa_2 \Lambda^{(k+2)}_{\alpha+2} (\Lambda^{(k+1)}_{\alpha+2})^{-1}
\eea
Since none of the elements of the diagonal $\Lambda$ matrices are zero,
\beq \kappa_1 \neq 0, \quad \kappa_2 \neq 0 \eeq
Similarly, on regeneration of systematic node $\alpha+2$, the components along $\alpha+1$ and $\alpha+3$ have to be aligned. Hence
\bea
\tilde{\kappa}_1 \Lambda^{(k+2)}_{\alpha+1} (\Lambda^{(k+1)}_{\alpha+1})^{-1} & = & \tilde{\kappa}_2 \Lambda^{(k+2)}_{\alpha+3} (\Lambda^{(k+1)}_{\alpha+3})^{-1} \label{eq:alpha3_lambdaSubstitute}
\eea
where $\tilde{\kappa}_1$ and $\tilde{\kappa}_2$ are some other non-zero constants in $\mathbb{F}_q$.

Now, for regeneration of systematic node $(\alpha+3)$ the component provided along it by the first non-systematic node is,
\begin{align}&\underline{x}^{(k+1,\alpha+3)} \Lambda^{(k+1)}_{\alpha+3} H_{\alpha+3} \nonumber \\ &\qquad=  \kappa_1 \underline{x}^{(k+2,\alpha+3)} \Lambda^{(k+2)}_{\alpha+1} (\Lambda^{(k+1)}_{\alpha+1})^{-1} \Lambda^{(k+1)}_{\alpha+3} H_{\alpha+3} \label{eq:alpha3_ind1} \end{align}
The right hand side of (\ref{eq:alpha3_ind1}) is obtained by substituting for $\underline{x}^{(k+1,\alpha+3)}$ from (\ref{eq:alpha3_xSubstitute}).

For regeneration of systematic node $(\alpha+3)$ the component provided along it by the second non-systematic node is,
\begin{align}&\underline{x}^{(k+2,\alpha+3)} \Lambda^{(k+2)}_{\alpha+3} H_{\alpha+3} \nonumber \\ & \qquad=  \tilde{\kappa_1}\tilde{\kappa_2}^{-1} \underline{x}^{(k+2,\alpha+3)} \Lambda^{(k+2)}_{\alpha+1} (\Lambda^{(k+1)}_{\alpha+1})^{-1} \Lambda^{(k+1)}_{\alpha+3} H_{\alpha+3} \label{eq:alpha3_ind2}\end{align}
The right hand side of (\ref{eq:alpha3_ind2}) is obtained by substituting for $\Lambda^{(k+2)}_{\alpha+3}$ from (\ref{eq:alpha3_lambdaSubstitute}).

From equations (\ref{eq:alpha3_ind1}) and (\ref{eq:alpha3_ind2}), it is clear that components along systematic node $(\alpha+3)$ node in the vectors passed by the two non-systematic nodes are linearly dependent. Hence by Lemma \ref{lem:indep} regeneration of node $(\alpha+3)$ node is not possible. \end{IEEEproof}

Since regeneration is not possible by connecting to $k-1$ systematic nodes and $\alpha$ non-systematic nodes, it will not be possible even in a general setting of using any $d$ nodes for regeneration.

\section{A Coding Scheme for any ($k,\alpha)$}\label{sec:simple_ach}
In this section, a coding scheme is described which can be used for any $(k, \alpha$) parameter set. This scheme assumes that when a systematic node fails, the existing $k-1$ systematic nodes and any $\alpha$ non-systematic nodes participate in the regeneration. This can be easily extended to a more general case.

\subsection{Scheme Description}
Divide the $k$ systematic nodes into $\alpha$ groups. Similar to the scheme given by Wu et al.\cite{WuDimISIT}, for regeneration of a systematic node, the existing systematic nodes in the same group as the failed node pass all their $\alpha$ symbols. The remaining systematic nodes and some $\alpha$ non-systematic nodes pass one symbol each.

The structure of the code is as follows. Let $\mu(l) \in \{1,\ldots,\alpha\}$ denote the group to which the systematic node $l$ belongs. Consider a set of variables $a_i^{(m)}$ and $b_{i,j}^{(m)}$, for $m=k+1,\ldots,n,\;i=1,\ldots,k,\;j=1,\ldots,\alpha,\;j \neq \mu(i)$. Let 
\beq \underline{b}_{i}^{(m)} = [b_{i,1}^{(m)}\cdots b_{i,\mu(i)-1}^{(m)}\; 0\;\; b_{i,\mu(i)+1}^{(m)}\cdots b_{i,\alpha}^{(m)}]\eeq
Let matrix $B_{i}^{(m)}$ be an $\alpha \times \alpha$ matrix such that it has $\underline{b}_{i}^{(m)}$ as its $\mu(i)^{th}$ row, and zeros elsewhere. Also let
\beq \tilde{\underline{b}}_{i}^{(m)} = [b_{i,1}^{(m)}\cdots b_{i,\mu(i)-1}^{(m)}\;\; a_i^{(m)}\;\; b_{i,\mu(i)+1}^{(m)}\cdots b_{i,\alpha}^{(m)}] \eeq

Let the node matrix of non-systematic node $m\;(\in \{k+1,\ldots,n\})$ be
\beq G_i^{(m)} = a_i^{(m)} I_{\alpha} + B_{i}^{(m)} \eeq
for $i=1,\ldots,k$, where $I_{\alpha}$ is an $\alpha \times \alpha$ identity matrix.

For example, suppose $k=5, \;\alpha=3$ and the systematic nodes are grouped as follows: \{1, 2\}, \{3\}, \{4, 5\}. Then, the node matrix stored by non-systematic node $m,\;( \in \{k+1,\ldots,n\})$ is
\beq \hspace{-7.7cm} \mathbf{G^{(m)}} = \eeq
\beq \hspace{-0.4cm} \left[ \resizebox{9cm}{!}{\begin{tabular}{*{3}{p{.5cm}}|*{3}{p{.5cm}}|*{3}{p{.5cm}}|*{3}{p{.5cm}}|*{3}{p{.5cm}}}
$a_1^{(m)}$&$b_{1,2}^{(m)}$&$b_{1,3}^{(m)}$&$a_2^{(m)}$&$b_{2,2}^{(m)}$&$b_{2,3}^{(1)}$&$a_3^{(m)}    $&$0         $&$0            $&$a_4^{(m)}    $&$0            $&$0        $&$a_5^{(m)}    $&$0            $&$0        $\\
$0        $&$a_1^{(m)}    $&$0            $&$0        $&$a_2^{(m)}    $&$0            $&$b_{3,1}^{(m)}$&$a_3^{(m)} $&$b_{3,3}^{(m)}$&$0            $&$a_4^{(m)}    $&$0        $&$0            $&$a_5^{(m)}    $&$0        $\\
$0        $&$0            $&$a_1^{(m)}    $&$0        $&$0            $&$a_2^{(m)}    $&$0            $&$ 0        $&$a_3^{(m)}    $&$b_{4,1}^{(m)}$&$b_{4,2}^{(m)}$&$a_4^{(m)}$&$b_{5,1}^{(m)}$&$b_{5,2}^{(m)}$&$a_5^{(m)}$\\
\multicolumn{6}{c}{$\underbrace{\qquad\qquad\qquad\qquad\qquad\qquad\qquad}$}&\multicolumn{3}{c}{$\underbrace{\qquad\qquad\qquad}$}&\multicolumn{6}{c}{$\underbrace{\qquad\qquad\qquad\qquad\qquad\qquad\qquad}$}\\ 
\multicolumn{6}{c}{\small{group 1}}&\multicolumn{3}{c}\small{group 2}\hspace{.8cm}&\multicolumn{6}{c}{\small{group 3}}
\vspace{-.6cm}
\end{tabular}} \right] \nonumber \eeq
\newline
\subsubsection{Regeneration}
Consider regeneration of systematic node $l\;(\in\{1,\ldots,k\})$. $\alpha$ non-systematic nodes, say $m_1,\ldots,m_{\alpha}$ pass the $\mu(l)^{th}$ row of their node matrices. The systematic nodes in other groups, say node $l^{\prime}$ in group $\mu(l^{\prime})\;\;(\mu(l^{\prime}) \neq \mu(l))$, pass the vector $[\underline{0}\cdots\underline{0}\;\underline{e}_{\mu(l)}\;\underline{0}\cdots\underline{0}]$ where the unit vector is in the position $l^{\prime}$. Since the component along node $l^{\prime}$ in the vector passed by any non-systematic node is $a_{l^{\prime}}^{(m)} \underline{e}_{\mu(l)}$, it can be subtracted out. The existing systematic nodes in group $\mu(l)$ pass all their symbols and hence components along these nodes can also be cancelled out. Hence, for regeneration, the components given out by the non-systematic nodes along the direction of the $l^{th}$ systematic node should be linearly independent. Thus, regeneration condition for systematic node $l$ with this choice of $\alpha$ non-systematic nodes reduces to a polynomial being non-zero i.e
\beq det \left(\begin{array}{c}
\tilde{\underline{b}}_{l}^{(m_1)}\\
\tilde{\underline{b}}_{l}^{(m_2)}\\
\vdots\\
\tilde{\underline{b}}_{l}^{(m_{\alpha})}\\
\end{array}\right)
\eeq
Similar polynomials are obtained $\forall l$, and for all sets of $\alpha$ non-systematic nodes. Clearly, none of these polynomials are identically zero. 

\subsubsection{Reconstruction}
If the data collector connects to the $k$ systematic nodes, then reconstruction is trivially satisfied. Consider DC connecting to $p$ non-systematic nodes, and $k-p$ systematic nodes, $1\leq p \leq k$. Let $m_1,\ldots,m_p,\quad(m_1<\ldots<m_p)$ be the non-systematic nodes to which it connects. Let $l_1,\ldots,l_p,\quad(l_1<\ldots<l_p)$ be the $p$ systematic nodes to which it does \textit{not}  connect. As in Section \ref{sec:recon}, reconstruction condition leads to following polynomial not equal to zero condition. 
\beq det \left(
\resizebox{3.8cm}{!}{\begin{tabular}{cccc}
$G_{l_1}^{(m_1)} $&$ G_{l_2}^{(m_1)} $&$ \cdots $&$ G_{l_p}^{(m_1)} $\\
$G_{l_1}^{(m_2)} $&$ G_{l_2}^{(m_2)} $&$ \cdots $&$ G_{l_p}^{(m_2)} $\\
$\vdots          $&$ \vdots          $&$        $&$ \vdots          $\\
$G_{l_1}^{(m_p)} $&$ G_{l_2}^{(m_p)} $&$ \cdots $&$ G_{l_p}^{(m_p)} $
\end{tabular}} \right)
\eeq

We will now show that there exists an assignment of the variables such that this polynomial is not identically zero. Set
\bea
\underline{b}_i^{(m)}&=&0 \quad \forall i,m\\
a_{l_i}^{(m_j)}&=& \left\lbrace \begin{array}{cc}
1 & \text{if } i=j\\
0 & \text{if } i \neq j
\end{array}
\right. \eea

By these assignments, the reconstruction matrix becomes an identity matrix, which is non-singular. Thus, the regeneration and reconstruction properties evaluate to the condition of the product of certain polynomials being non-zero. It is shown that none of these polynomials is identically zero. Assignment of values to the variables satisfying all the conditions can be obtained using the algorithm given by Koetter and Medard \cite{KotMed}.

This scheme can be extended to regeneration using any combination of systematic and non-systematic nodes provided that the systematic nodes in the same group as the failed node participate in regeneration. The extended proof will involve a few more conditions of polynomials being non-zero.

\subsection{Analysis}
\begin{figure}[t]
\vspace{-10pt}
  \centering
    \includegraphics[width=0.5\textwidth]{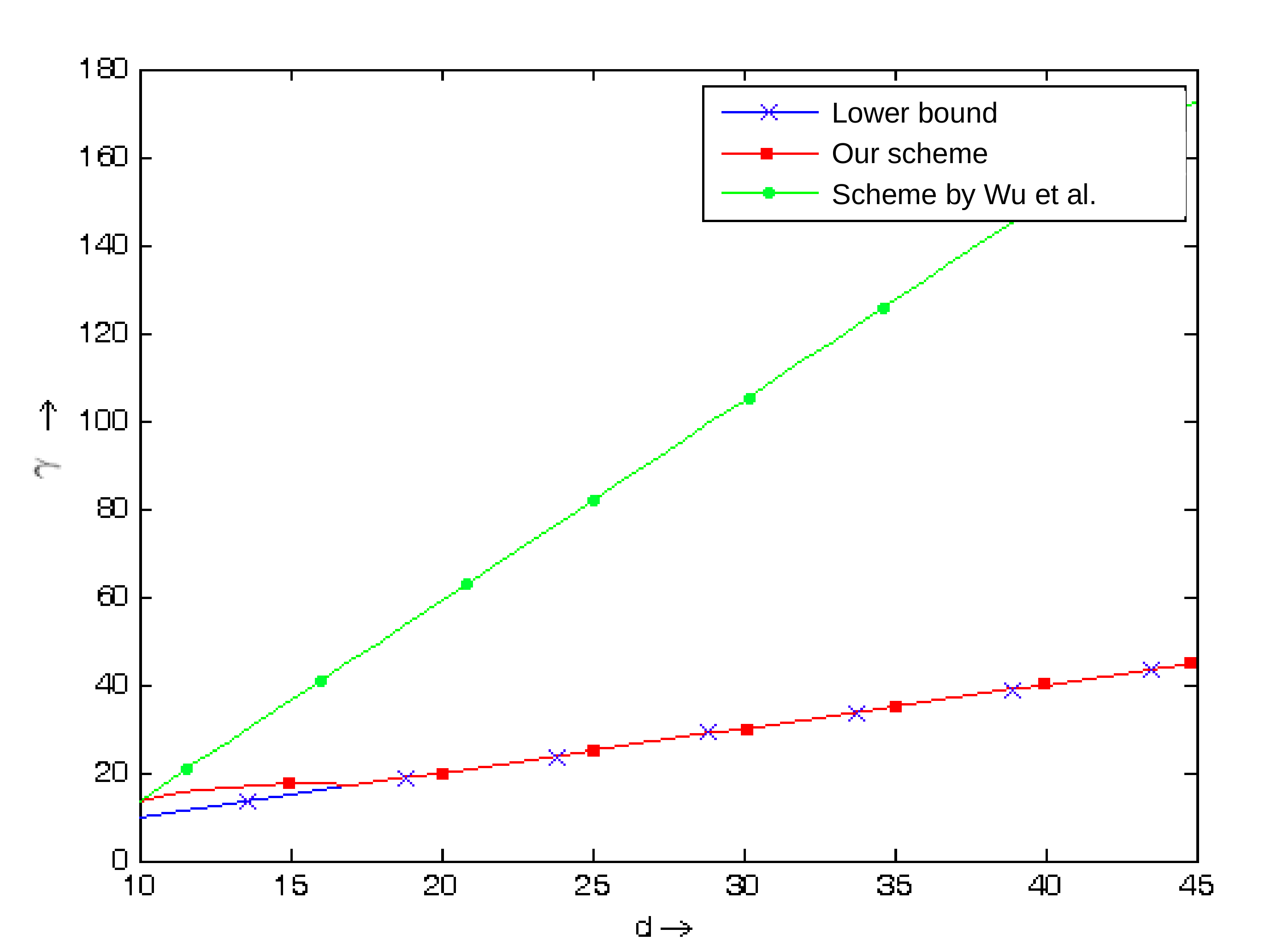}
    \vspace{-5pt}
\caption{Average repair bandwidth($\gamma$) required for exact regeneration of the systematic nodes with $\beta=1$ is plotted for various values of $d$ for $k=9$.} \label{fig:graph}
\end{figure}

For $k \leq \alpha$, if all the $\alpha$ nodes are kept in different groups, this scheme achieves the minimum repair bandwidth and hence is optimal. 

For $k > \alpha$, the amount of data to be downloaded for exact regeneration of a systematic node depends on the number of nodes in its group. If there are $\eta$ nodes in a group, the total number of symbols required to regenerate a node in that group,  is given by:
\beq \gamma = (\eta-1)\alpha + (d-\eta+1) \label{eq:gama}\eeq

\begin{lem}
The average repair bandwidth for exact regeneration of systematic nodes using the above described scheme is minimum when the groups are uniformly divided.
\end{lem}
\begin{IEEEproof}
Directly follows from equation (\ref{eq:gama})
\end{IEEEproof}

Let \beq s = \lfloor k / \alpha \rfloor\eeq 
Uniform division of groups would imply that out of the $\alpha$ groups, $k \bmod \alpha$ groups contain $s+1$ nodes each and the rest contain $s$ nodes each.

The average amount of download required for exact regeneration of the systematic nodes in our scheme is compared with the scheme proposed by Wu and Dimakis \cite{WuDimISIT} (\textit{Group interference alignment}) in Figure \ref{fig:graph}. The lower bound on the repair bandwidth is also plotted along side. It can be seen that for $d \geq 2k -1$ (i.e. $k \leq \alpha$) our scheme achieves the lower bound. For smaller values of $d$, the amount of data downloaded is higher. However, whether this achieved value of repair bandwidth is optimal or not is not known for $d \leq 2k-4$.

\begin{appendix}
Proof of Theorem \ref{thm:fullrank}

\begin{IEEEproof}
Let $\omega_1,\ldots,\omega_{k-p}$ ($\omega_1 <\ldots < \omega_{k-p}$) be the systematic nodes to which DC connects, and $\Omega_1,\ldots,\Omega_p$ ($\Omega_1 <\ldots < \Omega_p$) be the $p$ systematic nodes to which it does \textit{not}  connect. Thus the sets $\omega_1,\ldots,\omega_{k-p}$ and $\Omega_1,\ldots,\Omega_p$ are disjoint. Let $\delta_1,\ldots,\delta_p$ be the $p$ non-systematic nodes to which DC connects. The reconstruction matrix $R$ is given by
\bea
R=\begin{bmatrix}
   \mathbf{G'}^{(\delta_1)} \\
\vdots \\
   \mathbf{G'}^{(\delta_p)}
  \end{bmatrix} = \begin{bmatrix}
G^{(\delta_1)}_{\Omega_1} & G^{(\delta_1)}_{\Omega_2} & \cdots & G^{(\delta_1)}_{\Omega_p} \\
\vdots & \vdots & & \vdots \\
G^{(\delta_p)}_{\Omega_1} & G^{(\delta_p)}_{\Omega_2} & \cdots & G^{(\delta_p)}_{\Omega_p}
\end{bmatrix}
\eea

Group the $\Omega_1^{th}$ rows of $\mathbf{G'}^{(\delta_m)}\;(m=1,\ldots,p)$ as the first $p$ rows of a new matrix $R'$, then $\Omega_2^{th}$ rows as the next $p$ rows, and so on. Hence, row number $\Omega_i$ of $\mathbf{G'}^{(\delta_m)}$ becomes the row number $p \times (i-1) + m$ in $R'$. Below these, group the $\omega_1^{th}$ rows, then the $\omega_2^{th}$ and so on. Row number $\omega_i$ of $\mathbf{G'}^{(\delta_m)}$ becomes the row number $p^2 + p \times (i-1) + m$ in $R'$. Hence there are $\alpha$ groups with $p$ rows each in $R'$.

Let $S$ be an $p \times \alpha$ matrix with elements $[S]_{i,j} = \psi_j^{(\delta_i)}, \; i=1,\ldots,p,\; j=1,\ldots,\alpha$. Let $T_{a,b}$ be an $p \times \alpha$ matrix with its $b^{th}$ column as $[\psi_a^{(\delta_1)},\ldots,\psi_a^{(\delta_p)}]^t$, and rest of the elements zero. Thus, the $b^{th}$ column of $T_{a,b}$ is identical to the $a^{th}$ column of $S$.

The columns of $R'$ are grouped into $p$ groups of $\alpha$ columns each. Thus the matrix $R'$ can be viewed as a block matrix, with each block of size $p \times \alpha$, and the dimension of $R'$ being $\alpha \times p$ blocks.

Let $[R']_{(i,j)}$ represent the $(i,j)^{th}$ block of $R'$. For $i=1,\ldots,p,\;j=1,\ldots,p$ we get
\bea
[R']_{(i,j)}=\left \lbrace \begin{array}{ll}
                            \epsilon S\;\; &\text{if } i = j \\
T_{\Omega_j,\Omega_i} \;\; &\text{if } i \neq j
                           \end{array} \right.
\eea

For $i=p+1,\ldots,\alpha,\;j=1,\ldots,p$,
\bea [R']_{(i,j)} = T_{\Omega_j,\omega_{i-p}} \eea
Thus,
\beq R' = 
\left[  \begin{array}{ccccc}
\epsilon S & T_{\Omega_2,\Omega_1} & T_{\Omega_3,\Omega_1} & \cdots & T_{\Omega_p,\Omega_1} \\
T_{\Omega_1,\Omega_2} & \epsilon S & T_{\Omega_3,\Omega_2} & \cdots & T_{\Omega_p,\Omega_2} \\
\vdots & \vdots & \vdots &   & \vdots \\
T_{\Omega_1,\Omega_p} & T_{\Omega_2,\Omega_p} & T_{\Omega_3,\Omega_p} & \cdots & \epsilon S \\
\hline 
T_{\Omega_1,\omega_1} & T_{\Omega_2,\omega_1} & T_{\Omega_3,\omega_1} & \cdots & T_{\Omega_p,\omega_1} \\
\vdots & \vdots & \vdots &  & \vdots \\
T_{\Omega_1,\omega_{k-p}} & T_{\Omega_2,\omega_{k-p}} & T_{\Omega_3,\omega_{k-p}}  & \cdots & T_{\Omega_p,\omega_{k-p}}  \\
\end{array} \right] \eeq

Let $\tilde{S}$ be the $p \times p$ matrix formed by the columns $\Omega_1,\ldots,\Omega_p$ of $S$. As $\tilde{S}$ is a submatrix of Cauchy matrix $\Psi$, it is invertible. Let $\hat{I}$ be an $p \times \alpha$ matrix with columns $\omega_1,\dots,\omega_{k-p}$ having some arbitrary values (denoted by $\phi$), and the remaining $p$ columns put together forming an identity matrix. Let $\hat{E}_{a,b}$ be an $p \times \alpha$ matrix with the element at position  $(a,b)$ as $1$ and all other elements $0$.

Multiply each of the $\alpha$ groups of $p$ rows by $\tilde{S}^{-1}$. This will cause the following transformation in $R'$: $S$ will be replaced by $\hat{I}$ and $T_{\Omega_a,b}$ will be replaced by $\hat{E}_{a,b}$.

The resultant matrix will be of the form:
\beq \left[  \begin{array}{ccccc}
\epsilon \hat{I} & \hat{E}_{2,\Omega_1} & \hat{E}_{3,\Omega_1} & \cdots & \hat{E}_{p,\Omega_1} \\
\hat{E}_{1,\Omega_2} & \epsilon \hat{I} & \hat{E}_{3,\Omega_2} & \cdots & \hat{E}_{p,\Omega_2} \\
\vdots & \vdots & \vdots &  & \vdots \\
\hat{E}_{1,\Omega_p} & \hat{E}_{2,\Omega_p} & \hat{E}_{3,\Omega_p} & \cdots & \epsilon \hat{I} \\

\hline
\hat{E}_{1,\omega_1} & \hat{E}_{2,\omega_1} & \hat{E}_{3,\omega_1} & \cdots & \hat{E}_{p,\omega_1} \\
\vdots & \vdots & \vdots &   & \vdots \\
\hat{E}_{1,\omega_{k-p}} & \hat{E}_{2,\omega_{k-p}} & \hat{E}_{3,\omega_{k-p}}  & \cdots & \hat{E}_{p,\omega_{k-p}}  \\
\end{array} \right] \eeq

In the groups of rows $p+1,\ldots,\alpha$, every row has exactly one non-zero element. Hence these rows and the corresponding columns $(\omega_1,\ldots,\omega_{k-p})$ are independent of all others and can be eliminated. Note that all the $\phi$ elements are present only in these columns and hence the actual values of $\phi$ do not matter. The resultant matrix will be a $p^2 \times p^2$ matrix of the following form:

\beq \left[  \begin{array}{ccccc}
\epsilon I_{p} & E_{2,1} & E_{3,1} & \cdots & E_{p,1} \\
E_{1,2} & \epsilon I_{p} & E_{3,2} & \cdots & E_{p,2} \\
\vdots & \vdots & \vdots & \vdots & \vdots \\
E_{1,p} & E_{2,p} & E_{3,p}  & \cdots & \epsilon I_{p} \\
\end{array} \right] \eeq

\noindent where $I_{p}$ is a $p \times p$ identity matrix and $E_{a,b}$ is an $p \times p$ matrix with the element in the position $(a,b)$ as $1$ and all other elements $0$.

For $i=1,\ldots,p$, the $i^{th}$ row(column) of the $i^{th}$ row(column) group respectively contains exactly one non-zero element, and hence is linearly independent of all others. After eliminating these rows (and corresponding columns) the remaining matrix is rearranged by placing the $i^{th}$ row(column) of the $j^{th}$ group adjacent to the $j^{th}$ row(column) of the $i^{th}$ group to form:

\beq \left[  \begin{array}{ccccccc}
\epsilon & 1        & 0        & 0        & \cdots & 0        & 0       \\
1        & \epsilon & 0        & 0        & \cdots & 0        & 0       \\
0        & 0        & \epsilon & 1    & \cdots & 0        & 0       \\
0    & 0        & 1        & \epsilon & \cdots & 0        & 0       \\
\vdots & & \vdots & \vdots &  & \vdots & \vdots \\
0        & 0        & 0        & 0        & \cdots & \epsilon & 1    \\
0    & 0        & 0        & 0        & \cdots & 1        & \epsilon
\end{array} \right] \eeq

This is a block diagonal matrix, and since $\epsilon^2 \neq 1$, is full rank.
\end{IEEEproof}

In the example of $k=\alpha=3$ considered in section \ref{sec:eg_recon} when the data collector connected to the first systematic node, and the first two non-systematic nodes, we have $p=2$, $\omega_1 = 1$, $\Omega_1=2$, $\Omega_2=3$, $\delta_1=4$, $\delta_2=5$ and $\epsilon = 2$. Here, 
\(
\begin{array}{lll}
D_2 &=& R,\\
%\vspace{.1cm}
S &=& \left[  \begin{array}{lll}
\psi_1^{(4)} & \psi_2^{(4)} & \psi_3^{(4)} \\
\psi_1^{(5)} & \psi_2^{(5)} & \psi_3^{(5)} 
\end{array} \right] , \\

T_{\Omega_1,\Omega_2} &=& 
\left[  \begin{array}{lll}
0 & 0 & \psi_2^{(4)} \\
0 & 0 & \psi_2^{(5)} 
\end{array} \right], \\

\hat{I} &=& \left[  \begin{array}{lll}
\phi & 2 & 0 \\
\phi & 0 & 2 
\end{array} \right], \\

\hat{E}_{1,\Omega_2} &=& 
\left[  \begin{array}{lll}
0 & 0 & 1 \\
0 & 0 & 0 
\end{array} \right] \text{and}\\

E_{1,2} &=& 
\left[  \begin{array}{ll}
0 & 1 \\
0 & 0 
\end{array} \right].
\end{array}
\)
\end{appendix}

\end{document}